\documentclass[journal,12pt,onecolumn,draftclsnofoot,]{IEEEtran}

\ifCLASSINFOpdf
 \else
\fi

\usepackage{wrapfig}
\usepackage{comment}
\usepackage{balance}  
\usepackage{cite}
\usepackage{graphicx}
\usepackage{float}
\usepackage{algorithm}
\usepackage{algpseudocode}
\usepackage{amsmath} 
\usepackage[utf8]{inputenc}
\usepackage[final]{pdfpages} 
\usepackage{pdfpages}
\usepackage{lipsum}
\usepackage{textcase}
\usepackage{url}
\usepackage{amsmath,esint}
\usepackage{epstopdf}
\usepackage{array} 
\usepackage{hyperref}
\usepackage{multirow}
\usepackage{subcaption}
\usepackage{booktabs}
\usepackage{esvect}
\usepackage{soul}
\usepackage{ulem}
\usepackage[scientific-notation=true]{siunitx}

\newcolumntype{P}[1]{>{\centering\arraybackslash}p{#1}}
\newcolumntype{M}[1]{>{\centering\arraybackslash}m{#1}}
\newcommand{\uvec}[1]{\boldsymbol{\hat{\textbf{#1}}}}

\begin{document}
%

\title{Doors in the Sky: Detection, Localization and Classification of Aerial Vehicles using Laser Mesh}

\author{
\IEEEauthorblockN{Wahab Khawaja, Ender Ozturk, and 
Ismail Guvenc} \IEEEmembership{Fellow, IEEE}\\
\thanks{This work was supported by NASA under the award NNX17AJ94A. W. Khawaja is 
with the Mirpur University of Science \& Technology, Pakistan. The co-authors are with NC State University, Raleigh, NC 27606 USA (e-mails: wahab.ali@must.edu.pk, \{wkhawaj, eozturk2, iguvenc\}@ncsu.edu). }
}


\maketitle
\begin{abstract}
\textit{The stealth technology and unmanned aerial vehicles~(UAVs) are expected to dominate current and future aerial warfare. The radar systems at their maximum operating ranges, however, are not always able to detect stealth and small UAVs mainly due to their small radar cross-sections and/or low altitudes. In this paper, a novel technique as an alternative to radar technology is proposed. The proposed approach is based on creating a mesh structure of laser beams initiated from aerial platforms towards the ground. The laser mesh acts as a virtual net in the sky. Any aerial vehicle disrupting the path of the laser beams are detected and subsequently localized and tracked. As an additional feature, steering of the beams can be used for increased coverage and improved localization and classification performance. A database of different types of aerial vehicles is created artificially based on Gaussian distributions. The database is used to develop several machine learning (ML) models using different algorithms to classify a target. Overall, we demonstrated through simulations that our proposed model achieves simultaneous detection, classification, localization, and tracking of a target.}
\end{abstract}


\IEEEpeerreviewmaketitle

\section{Introduction} 
Unmanned aerial vehicles~(UAVs) have applications in several areas nowadays~\cite{wahab_survey}, one of which is the defense industry. The small size and ability to fly at low altitudes make the UAVs practically invisible to conventional radar systems~\cite{radar_uav} at long ranges. Moreover, stealth technology is the most essential part of the current and future combat aerial vehicles. The stealth aerial vehicles~(SAVs) can avoid being detected by radars due to their small radar cross-sections~(RCS) that is achieved by using a mix of techniques to absorb and scatter the incoming radar energy~\cite{stealth}. 

Radar technology has evolved significantly over the decades. However, there are still known limitations of the radar systems~\cite{radar_limit}. The basic radar principle still relies on back-scattered reflections from a potential target~\cite{radar_basic}. For SAVs and UAVs, the back-scattered reflections are weak and the strength of the received signal is generally below the noise floor. Therefore, the target remains undetected by conventional radar systems over a significant distance, i.e., insufficient slant range. In the literature, there are some radar systems proposed for detecting targets with small RCS~\cite{antistealth,antistealth2,antistealth3}. However, these radar systems are too complex, expensive, and they are designed to use only specific wavelengths that further contributes to complexity and high cost.

Different types of early warning radar systems can provide specific detection capabilities against the SAVs and UAVs~\cite{early_warning1,early_warning2}. However, the detection capabilities~(i.e, RCS as a function of detection range) vary with the operating frequency of the radar system and the type of the aerial vehicle. There are also laser scanning techniques available in the literature for the detection and tracking of terrestrial moving objects~\cite{laser1,laser2,laser3}. The laser scanning works on a similar principle to radars, i.e., relies on back-scattered signal, and classifies a moving target based on individual laser scans at different instances of the time obtained from different parts of a target. Implementation of this technique is quite challenging, and there is limited amount of work available in the literature for the detection of aerial targets using this technique. In \cite{laserUAV}, laser-based radar is used for the detection of UAVs. As the approach in \cite{laserUAV} follows the basic radar principle, it also shares the disadvantages of the radar systems.     

Similar to radar systems, electro-optical/Infra-red~(EO/IR) imaging is used for detection, tracking, and classification of aerial targets~\cite{EOIR}. The operation of the EO/IR sensors are different from the radar systems. The EO/IR imaging uses ultraviolet, visible and infrared spectral bands for different types of targets. The advantage of EO/IR technique compared to radars is that it can operate in the passive mode and the illumination is either provided by natural sources or by the target itself. The fine details of the target filtered from the background environment using EO/IR image processing can help in the tracking, and classification of small and stealthy targets. However, the selection of the spectral band for EO/IR imaging is dependent on the type of the target. The EO/IR imaging is also affected by environmental and atmospheric conditions, e.g. haze, fog, and clouds. The EO/IR imaging has limited range compared to radars. The high resolution and sensitivity EO/IR imaging is complex and expensive.  

\begin{figure*}
\centering
\includegraphics[width=0.7\textwidth]{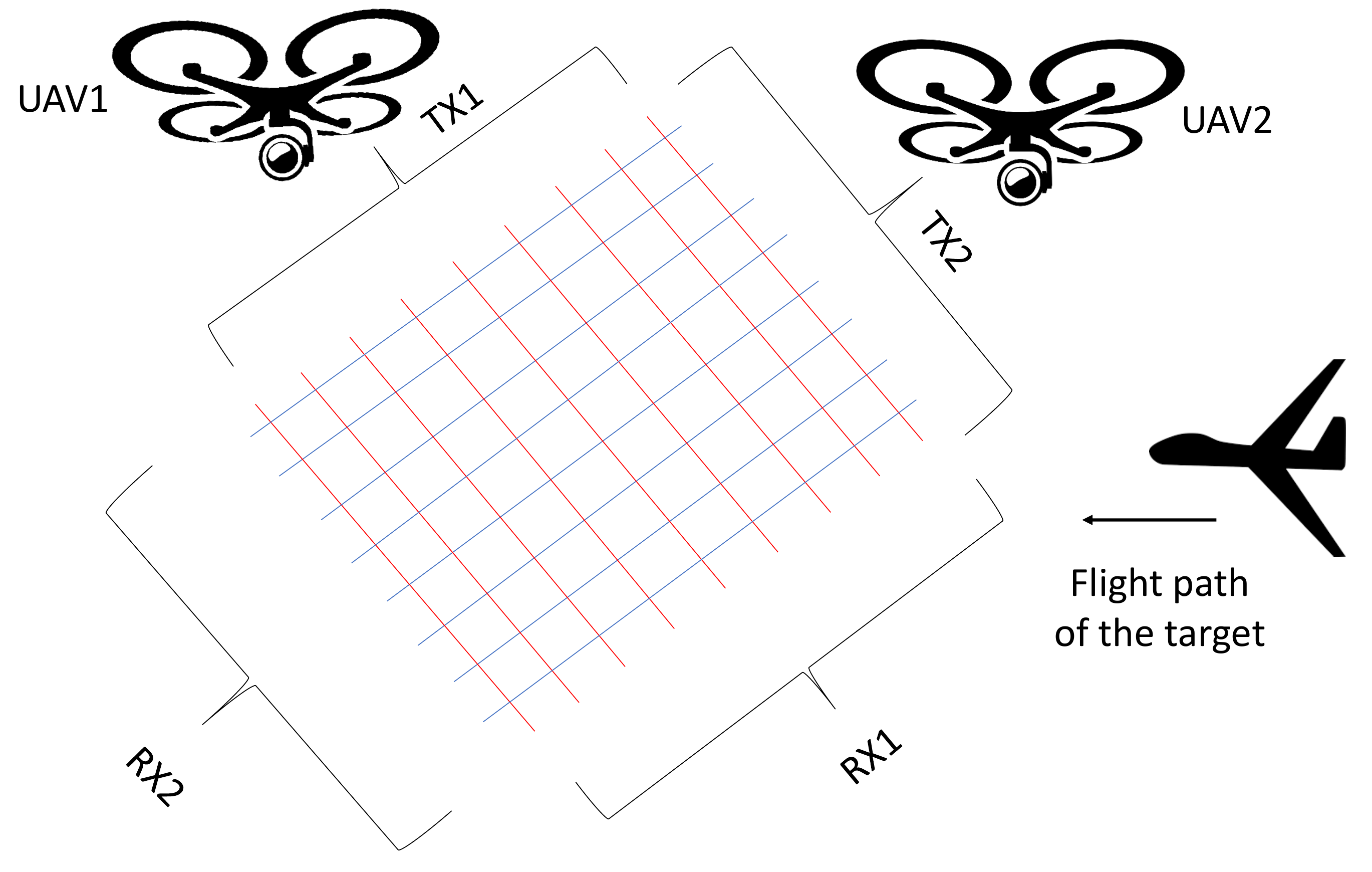}
\caption{Laser mesh created by two airborne UAVs that are quasi-static. A uniformly spaced laser array is considered on each airborne UAV. The laser array platform is tilted at each airborne UAV. The coverage area and shape of the elements of the mesh can be changed by changing the spacing between the individual laser beams and changing the position of airborne UAVs or changing the tilt of the array platform.} \label{Fig:Laser_mesh} 
\end{figure*}

In this paper, a new alternative to radar systems for the detection, classification, localization, and tracking of aerial objects is provided. In our work, a novel concept of the mesh of laser beams in the sky is proposed. A simple illustration of the concept using two UAVs is shown in Fig.~\ref{Fig:Laser_mesh}. The laser mesh forms a web-like structure in the sky. Laser beams in the mesh can be transmitted either from a satellite or a high altitude platform~(HAP) such as hot air balloons, medium altitude rotary-wing UAVs, or both.  The transmitted laser beams can be received either directly on the ground or rotary-wing UAVs near the ground. Any flying object that crosses the laser mesh will block the path of the laser beams and will be detected and subsequently localized. Laser mesh steering is introduced to steer virtual gates at different azimuth positions. The laser steering in the azimuth plane helps in localization, classification, and tracking of moving aerial objects, and increases the coverage area. A mathematical model of the blockage of the laser beam of the Gaussian profile is provided. Moreover, Gaussian training data that has the features of 3D shape, maximum velocity, pitch and drift angles, and a maximum altitude is used for model development. Dataset is created for eleven different type of flying objects grouped in four categories. The results obtained from simulations proves the viability of our proposed approach. 

The main advantages of our proposed approach compared to radar systems are summarized as follows:
\begin{itemize}
\item Our proposed approach is energy efficient compared to the radar systems. The radar systems broadcast the radio signals in the free space and the transmitted energy spreads in different directions. A small fraction of the transmitted energy from the radar is received back. On the other hand, in our approach, a point-to-point connection is established without significant spreading of the transmitted energy towards the receiver~(RX). Moreover, the received laser energy can be recycled into other forms. 
\item The energy emissions from the radar can be detected from a long distance, hence, exposes them to the risk of detection. However, in the proposed approach, there are no long-range emissions sourced from the transmitter~(TX) or the RX.
\item Radar systems use complex algorithms for clutter rejection. These algorithms have limitations dependent on the size and height of the flying target, the motion of the objects in the surrounding, and the terrain type. In the proposed approach, no clutter rejection is required. 
\item The radar systems depend on the delay of the received pulses to range a target and pulse repetition frequency needs to be correctly selected to avoid range ambiguity. In our approach, a target is detected instantaneously without delay and range ambiguity issue. 
\item Large and bulky antennas are required for long-range transmission and reception of radar signals. There are additional maintenance overheads for rotating radars. On the other hand, laser beam generation, and reception are performed using electronic and optical equipment that are concise and more efficient compared to mechanically controlled radar systems. 
\item Complex sounding signals and processing techniques are used for the detection, localization, and classification of sophisticated targets e.g., terrain hugging drones and missiles. Our proposed approach does not require complex sounding signals and associated processing techniques. 
\item The probability of detection of different types of aerial targets with the proposed approach is significantly higher compared to the conventional radar systems. The probability of false alarms using our approach is also small compared to radar systems as the detection is solely dependent on the interruption of the laser link. 
\item We can achieve simultaneous detection, classification, localization, and tracking~(SDCLT) using our approach with high accuracy and without any significant complexity. For radar systems, SDCLT requires complex operations and accuracy depends on the reflected energy that can be manipulated by the target.
\item The common countermeasure for engaging incoming aerial targets mainly consists of firing projectiles towards the incoming aerial object. The interception procedure is highly complicated and requires constant tracking of the aerial object from multiple sources. However, using the proposed setup, the interception procedure can be significantly simplified.    
\item The range resolution of traditional radar systems has a trade-off with the detection range~\cite{range_amb}. Similarly, the angular resolution has a trade-off with the instantaneous field of view of the radar system. On the other hand, for our proposed approach, the resolution of the target depends only on the density of the laser beams.   
\item The main information of a target obtained by a radar system is its RCS. The accurate determination of the RCS of a target is important for the classification of modern aerial threats. The RCS of a target depends on many factors e.g., frequency of the radar system, angle of illumination, and physical properties of the target. Therefore, the RCS of a target varies during measurements. No RCS variations are present using our approach.  
\item The RXs of modern radar systems are complex and expensive. The complexity is mainly due to processing of the 1) weak received echoes, 2) range and Doppler ambiguities, 3) clutter rejection, and 4) finding the precise position of the target in the azimuth and elevation planes. In comparison, no complex processing is required at the RXs of our approach. The RX components of our proposed approach are also simple and inexpensive. 
\item Generally, a single radar cannot perform all the tasks related to an aerial target. For example, there are separate search, tracking, and fire-control radar systems. These radar systems require networking for a combined response. The networking is vulnerable to failures, jamming, and delays. Our proposed laser mesh setup provides a single front-end for detection, tracking, classification, and fire-control guidance. 
\item Radar Systems are more vulnerable to mechanical and electronic jamming compared to our proposed approach.
\item The performance of a particular radar system depends on the terrain. For example, the performance of a particular radar system is different in a hilly area compared to an urban area. In comparison, our proposed approach is independent of the terrain.
\end{itemize}

\begin{figure*}
\centering
\includegraphics[width=0.6\textwidth]{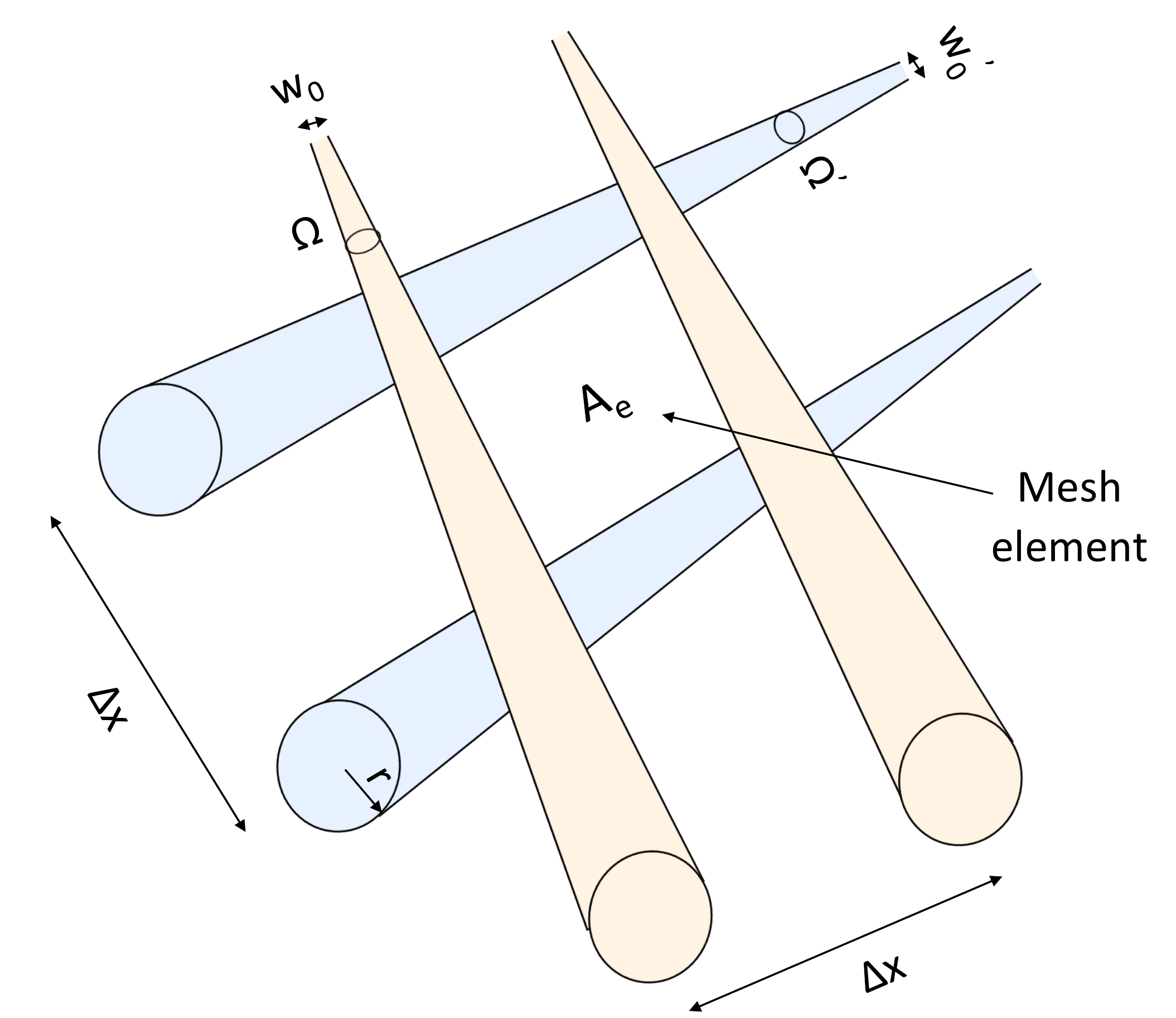}
\caption{A mesh element is created by laser beams from two sources. The divergence of the laser beams along the direction of propagation is also shown. The divergence of the laser beams is considered small resulting in approximately the same beam radius along the direction of propagation. }\label{Fig:Laser_mesh_new} 
\end{figure*}

The rest of the paper is organized as follows: The proposed laser mesh setup is provided in Section~\ref{Section:Propagation_Scenario}, Section~\ref{Section:Detection} discusses the detection of a target, the classification features of a target are provided in Section~\ref{Section:Features}, Section~\ref{Section:Classification} provides classification, localization, and tracking of a target, the limitations of the proposed approach are discussed in Section~\ref{Section:limitations}, the simulation results are provided in Section~\ref{Section:Simulation_Results}, and Section~\ref{Section:Conclusions} concludes the paper. 

\section{Proposed Laser Mesh Setup} \label{Section:Propagation_Scenario}
In this section, the details of the laser mesh setup and laser mesh steering are provided.

\subsection{Mesh of Laser Beams} \label{Section:laser_mesh}

Laser is a concentrated beam of light obtained through stimulated emission of electromagnetic radiation~\cite{laser_ref}. The main advantage of a laser is spatial coherence resulting in higher directivity compared to radio waves. A single laser beam has a small coverage area due to high directivity. Therefore, to cover a large area, multiple laser beams are required. The coverage area of a single laser beam at a given distance from the TX depends on the divergence of the beam. The divergence of the laser beam with distance results in the broadening of the beam, as shown in Fig.~\ref{Fig:Laser_mesh_new}. We consider that the divergence of the laser beams is small~(e.g. collimated laser beams). The small divergence of laser beam results in approximately constant radius of the beam with distance. In Fig.~\ref{Fig:Laser_mesh_new}, $\Omega$ and $\Omega'$, are the solid angles of the beams, and $w_0$ and $w_0'$ are the beam waists of the two sources. Given $r$ as the radius of the beam, and $\Delta x$ as the separation between the two beams such that $\Delta x\gg r$, the area of a mesh element in Fig.~\ref{Fig:Laser_mesh_new} is approximated as $A_{\rm e} = (\Delta x)^2$~\cite{beam_div}. The area of the laser mesh element, $A_{\rm e}$, is selected depending on the type of the aerial threat. 

In our proposed approach, the TXs~(and RXs) of the laser beams can be either on a satellite, a HAP, or on a medium altitude hovering UAV. The RXs~(or TXs) can be on the ground or over a hovering UAV. Multiple laser beams can be transmitted from a single TX. A blockage to the path of a laser beam by any aerial object~(taken as target) is readily identified. Localizing a target in both azimuth and elevation planes based on the blockage to the path of the laser beams by the target will require at least two sources of laser beams as shown in Fig.~\ref{Fig:Laser_mesh}. The laser beams from two different sources in a laser mesh can be overlapping or non-overlapping. Different wavelength lasers will be needed to avoid co-channel interference in case of overlapping beams.

\subsection{Laser Mesh Steering} \label{Section:steer}
\begin{figure*}[!t]
	\centering
	\includegraphics[width=0.8\textwidth]{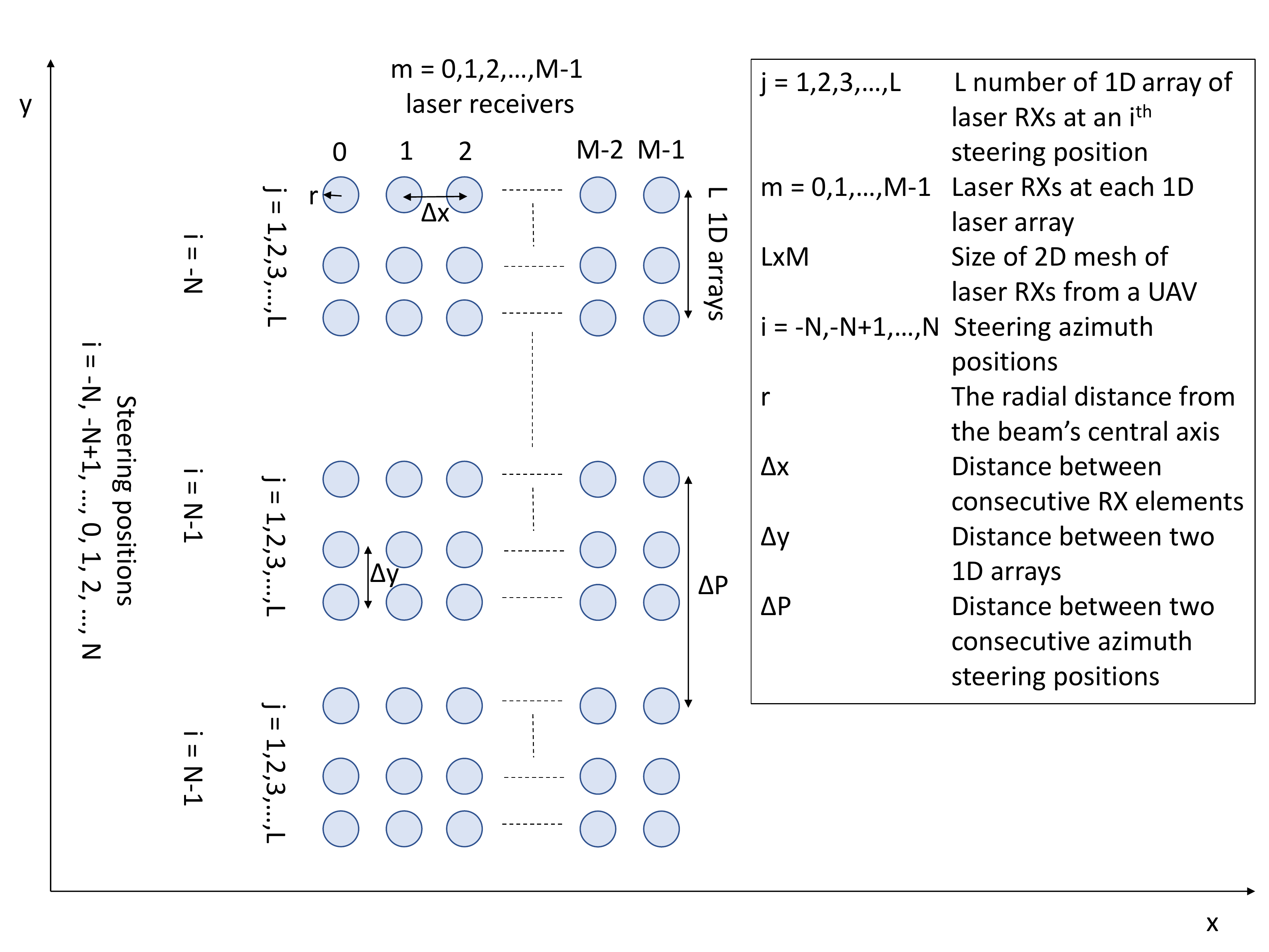} 
	\caption{Top view of the RXs on the ground for laser beams from a UAV (e.g. UAV1 in Fig.~\ref{Fig:Laser_mesh}). Laser mesh steering is used to steer the transmitted laser beams to different RXs positions on the ground. There are $2N+1$ steering positions, and each steering position has $L$ number of 1D array of laser RXs. The number of laser beam RXs in each 1D array is $M$. } \label{Fig:Laser_steer} 
\end{figure*}

The coverage area of the laser mesh, i.e., the size of the net, in the azimuth plane depends on the number and separation distances of the laser RXs. We will assume that the laser RXs are placed on the ground. The top view of the steered laser positions~(RXs) in the $(x,y)$ plane is given in Fig.~\ref{Fig:Laser_steer}. In this figure, a single two dimensional~(2D) laser mesh from an airborne UAV~(e.g. UAV1 in Fig.~\ref{Fig:Laser_mesh}) is composed of $L\times M$ laser RXs at $i^{\rm th}$ steering position. The 2D mesh contains $j=1,2,3,\hdots,L$, one dimensional~(1D) array of laser RXs, and $M$ is the number of RX elements at each array, shown in Fig.~\ref{Fig:Laser_steer}. 

In Fig.~\ref{Fig:Laser_steer}, there are $2N+1$ azimuth positions due to beam steering. The steering centers at the azimuth positions are $i=-N,-N+1,\hdots,0,1,2,\hdots,N$. The distance between the consecutive RXs is represented as $\Delta x$, the distance between two consecutive 1D arrays is represented as $\Delta y$ and the distance between the two consecutive steering positions is $\Delta P$. The $\Delta x$ and $\Delta y$ should be carefully chosen to classify the aerial targets based on their dimensions. From Fig.~\ref{Fig:Laser_steer}, a matrix of dimensions $L(2N+1)\times M$ is obtained over all the steering positions.

If $h$ is the height of the TX from the RXs at the central steering position, $i=0$, $j = L/2$, the slant range, $d_{i,j}$ from the TX to $i^{\rm th}$ steering position~($i\neq~0$) and $j^{\rm th}$ 1D array is $d_{i,j} = \sqrt{h^2 + \big( (j - L/2)\Delta y + i \Delta P\big)^2}$. The total azimuth distance covered during steering is $2N\Delta P L\Delta y$. In addition, the speed of the steering between any two consecutive steering positions is given as $v_{\rm s} =\frac{\Delta P}{\Delta t_{\rm s}}$, where $\Delta t_{\rm s}$ is the time to steer between consecutive steering positions. If $v$ is the maximum speed of the incoming target, then it is required that $v_{\rm s}\gg v$ in order to accurately detect and localize the moving object.

\section{Detection of a Target} \label{Section:Detection}
In this section, a Gaussian laser beam is considered and the blockage to the path of a Gaussian beam is mathematically modeled. The model is then used to detect the presence of a target.

\subsection{Mathematical Modeling of Blockage of a Laser beam} \label{Section:Laser_mesh_new}
\begin{figure*}[!t]
	\centering
	\includegraphics[width=0.60\textwidth]{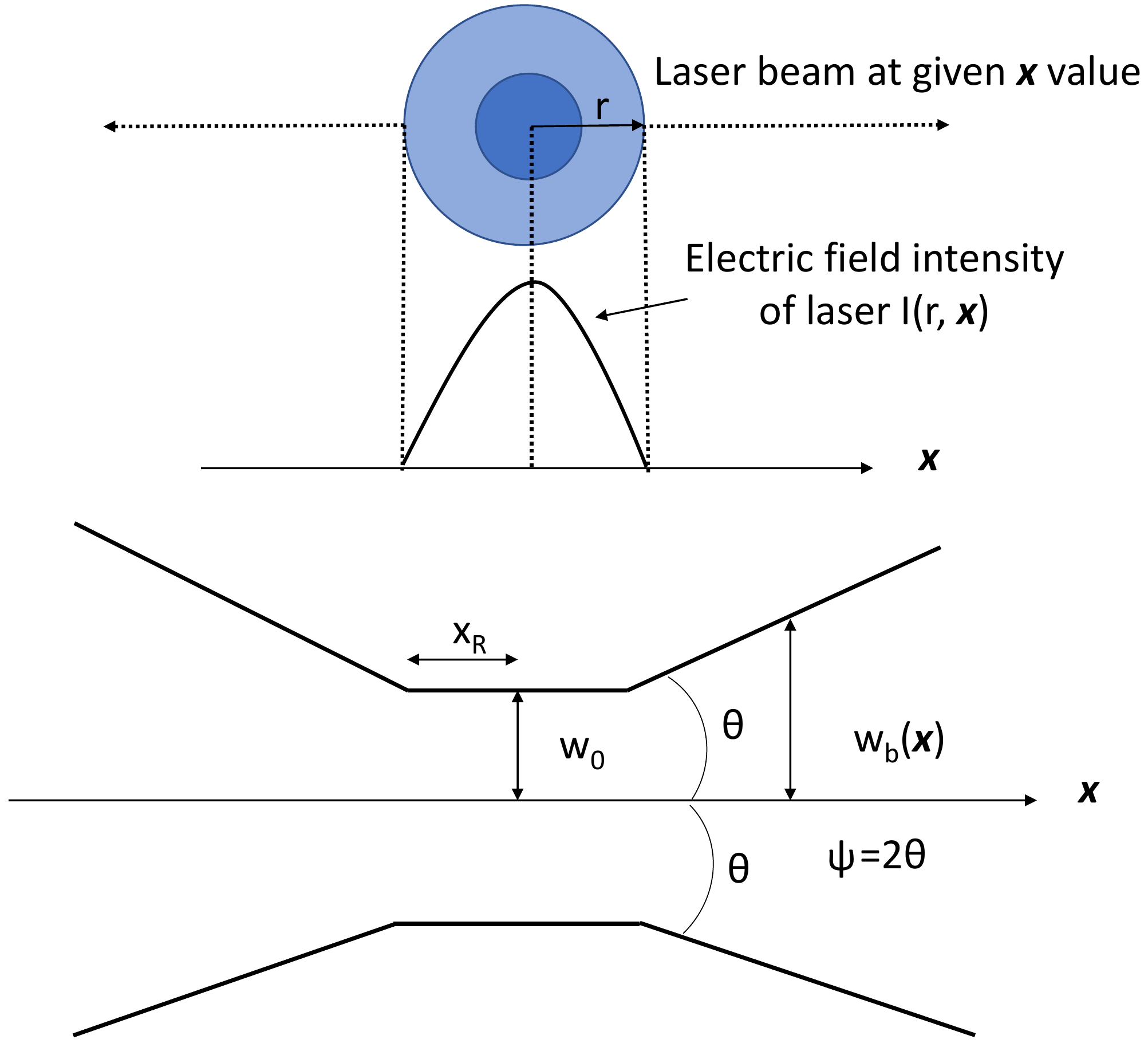} 
	\caption{A Gaussian profile of a single laser beam propagating in the $x$-direction.} \label{Fig:Guassian_beam} 
\end{figure*}

Laser beams considered in this work are modeled as Gaussian beams~\cite{laser11}. Consider a single Gaussian beam illustrated in Fig.~\ref{Fig:Guassian_beam}. The direction of the propagation of the beam is along the $x$-direction and the beam is polarized in the $z$-direction. The incident electric field $\textbf{E}^{\rm (inc)}(r,x)$ for the beam at a distance $x$ from the source, using paraxial approximation, given as~\cite{laser22}
\begin{align}
    &\textbf{E}^{\rm (inc)}(r,x) = \nonumber \\  & \label{Eq:E_inc} E_0 \uvec{z} \frac{w_0}{w_{\rm b}(x)}\exp\bigg(\frac{-r^2}{w_{\rm b}^2(x)}\bigg)\exp\bigg(-j(k x + \frac{k r^2}{2R(x)} - \phi(x) )\bigg),  
\end{align}
where $r$ represents the radial distance from the beam's central axis, $E_0$ is the electric field amplitude at origin and at time instance~$t_0=0$. $w_{\rm b}(x)$ is the width of the beam along the direction of propagation, which is given as~\cite{laser22}
\begin{align}
    w_{\rm b}(x) = w_0\sqrt{1+\frac{x}{x_R}},
\end{align}
where $x_R$ is the Rayleigh range given by $x_R = \frac{\pi w_0^2 n_0}{\lambda}$, $\lambda$ is the wavelength, and $n_0$ is the index of refraction of the free space. In \ref{Eq:E_inc}, $k = \frac{2 \pi n_0}{\lambda}$ is the wavenumber, $R(x)$ is the radius of curvature of the wavefront of the beam at an axial distance $x$, given by $R(x) = x\bigg[1 + (\frac{x_R}{x})^2\bigg]$, and $\phi(x)$ is the Gouy phase. The divergence of the laser beam in Fig.~\ref{Fig:Guassian_beam} is represented by angle $\theta$ for $x \gg x_R$ as $\displaystyle{\theta = \lim_{x \to \infty}\arctan\big(\frac{w_{\rm b}(x)}{x}}\big)$. Similarly, the apex angle of the cone is given as $\psi = 2\theta$, and solid angle $\Omega = \pi \sin^2\theta$. Moreover, the incident magnetic field $\textbf{H}^{\rm inc}(r,x)$ polarized in the $y$-direction is given as $\textbf{H}^{\rm (inc)}(r,x) = \frac{\uvec{y}}{\eta_0}{E}^{\rm (inc)}(r,x)$, where $\eta_0$ is the impedance of the free space. 

The incident intensity distribution is given by
\begin{align}
    I^{(\rm inc)}(r,x) = &\frac{{\rm Re}\bigg(E^{(\rm inc)}\times H^{*(\rm inc)}\bigg)}{2}, \nonumber \\
    =&\frac{|E_0|^2}{2\eta_0}\bigg(\frac{w_0}{w_{\rm b}(x)}\bigg)^2\exp\bigg(\frac{-2r^2}{w_{\rm b}^2(x)}\bigg), \label{Eq:Inten_inc}
\end{align}
where $\frac{|E_0|^2}{2\eta_0}$ is the intensity at the beam's waist. In the case of blockage, the incident electric field is divided into reflected and transmitted electric fields given as $\textbf{E}^{\rm (rfl)}(r,x)$, and $\textbf{E}^{\rm (tx)}(r,x)$, respectively. Therefore, the incident electric field can be written as $\textbf{E}^{\rm (inc)}(r,x) = -\textbf{E}^{\rm (rfl)}(r,x) + \textbf{E}^{\rm (tx)}(r,x)$. The reflected component, $\textbf{E}^{\rm (rfl)}(r,x)$ is mainly specular as the wavelength is significantly small compared to the size of the target. The transmitted component is significantly small compared to the reflected component for solid surface targets. The reflected and transmitted intensity distributions represented, respectively, as $I^{(\rm rfl)}(r,x)$ and $I^{(\rm tx)}(r,x)$ are 
\begin{align}
    I^{(\rm rfl)}(r,x) &= \Gamma_1^2I^{(\rm inc)}(r,x), \nonumber \\
    I^{(\rm tx)}(r,x) &= \Gamma_2^2I^{(\rm inc)}(r,x), \label{Eq:inten}
\end{align}
where $\Gamma_1 = \frac{\big|E^{\rm (rfl)}(r,x)\big|}{\big|E^{\rm (inc)}(r,x)\big|}$ is the reflection coefficient, and $\Gamma_2 = \frac{\big|\textbf{E}^{\rm (tx)}(r,x)\big|}{\big|\textbf{E}^{\rm (inc)}(r,x)\big|}$ is the transmission coefficient.

\subsection{Detection of the Target based on Blockage of Laser Beam} \label{Section:Target_detection}
The RXs of laser beams at $i^{\rm th}$ azimuth steering position and $j^{\rm th}$ mesh, at a distance $d_{i,j}$ from the TX, and an aperture radius of $r^{\rm (RX)}$ expects an intensity $I^{(\rm RX)}(r^{\rm (RX)},d_{i,j})$ \big(See \ref{Eq:Inten_inc} \big) in case there is no blockage. The intensity, $I^{(\rm RX)}$, as a function of $r^{\rm (RX)}$ and $d_{i,j})$ is given as 
\begin{align}
 I^{(\rm RX)} (r^{\rm (RX)},d_{i,j})= \frac{|E_0|^2}{2\eta_0}\bigg(\frac{w_0}{w_{\rm b}(d_{i,j})}\bigg)^2\exp\bigg(\frac{-2r{^{(\rm RX)}}^2}{w_{\rm b}^2(d_{i,j})}\bigg). \label{Eq:rx_inten}
\end{align}
The corresponding received power at the RX side is given as~\cite{laser33} 
\begin{align}
    P^{\rm (RX)}(r^{(\rm RX)},d_{i,j}) = \frac{\pi\big|E_0|^2w_0^2\bigg[1 - \exp \bigg(\frac{-2r{^{(\rm RX)}}^2}{w_b^2(d_{i,j})}\bigg)\bigg]}{4\eta_0}. \label{Eq:RX_pwr}
\end{align}
Finally, the signal-to-noise ratio~(SNR) represented as $S/n_{\rm G}$ at the RXs of $i^{\rm th}$ steering position and $j^{\rm th}$ mesh in the presence of additive white Gaussian noise~(AWGN) $n_{\rm G} \sim \mathcal{N}(0,\,\sigma^{2}_{\rm n})\,$ can be written as 
\begin{align}
   S/n_{\rm G} = \frac{ P^{\rm (RX)}(r^{(\rm RX)},d_{i,j})}{\sigma^{2}_{n}}. \label{Eq:SNR}
\end{align}

\begin{figure*}[!t]
	\begin{subfigure}{0.56\columnwidth}
	\centering
	\includegraphics[width=\textwidth]{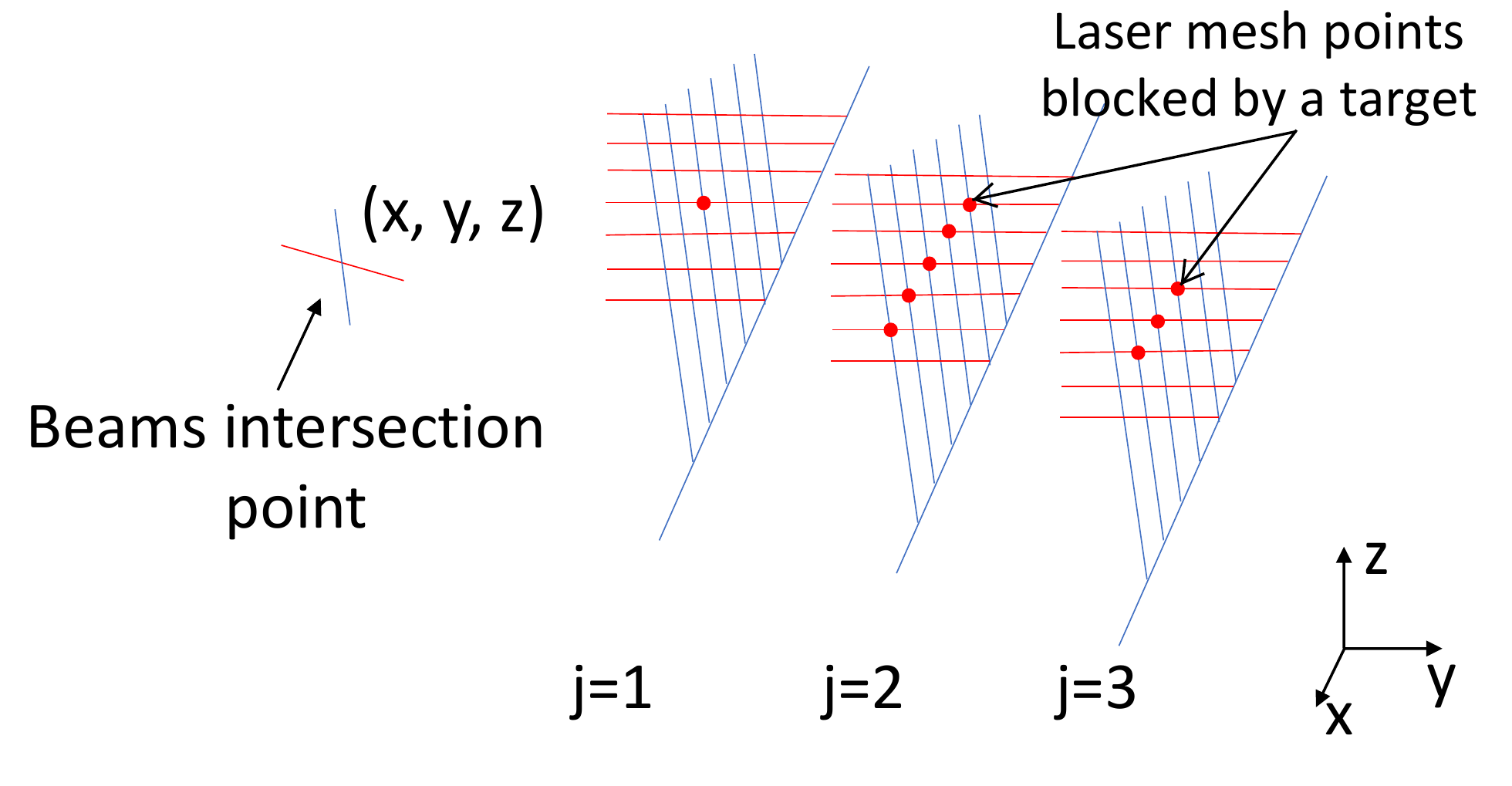} 
	\caption{}
    \end{subfigure}			
	\begin{subfigure}{0.48\columnwidth}
	\centering
    \includegraphics[width=\textwidth]{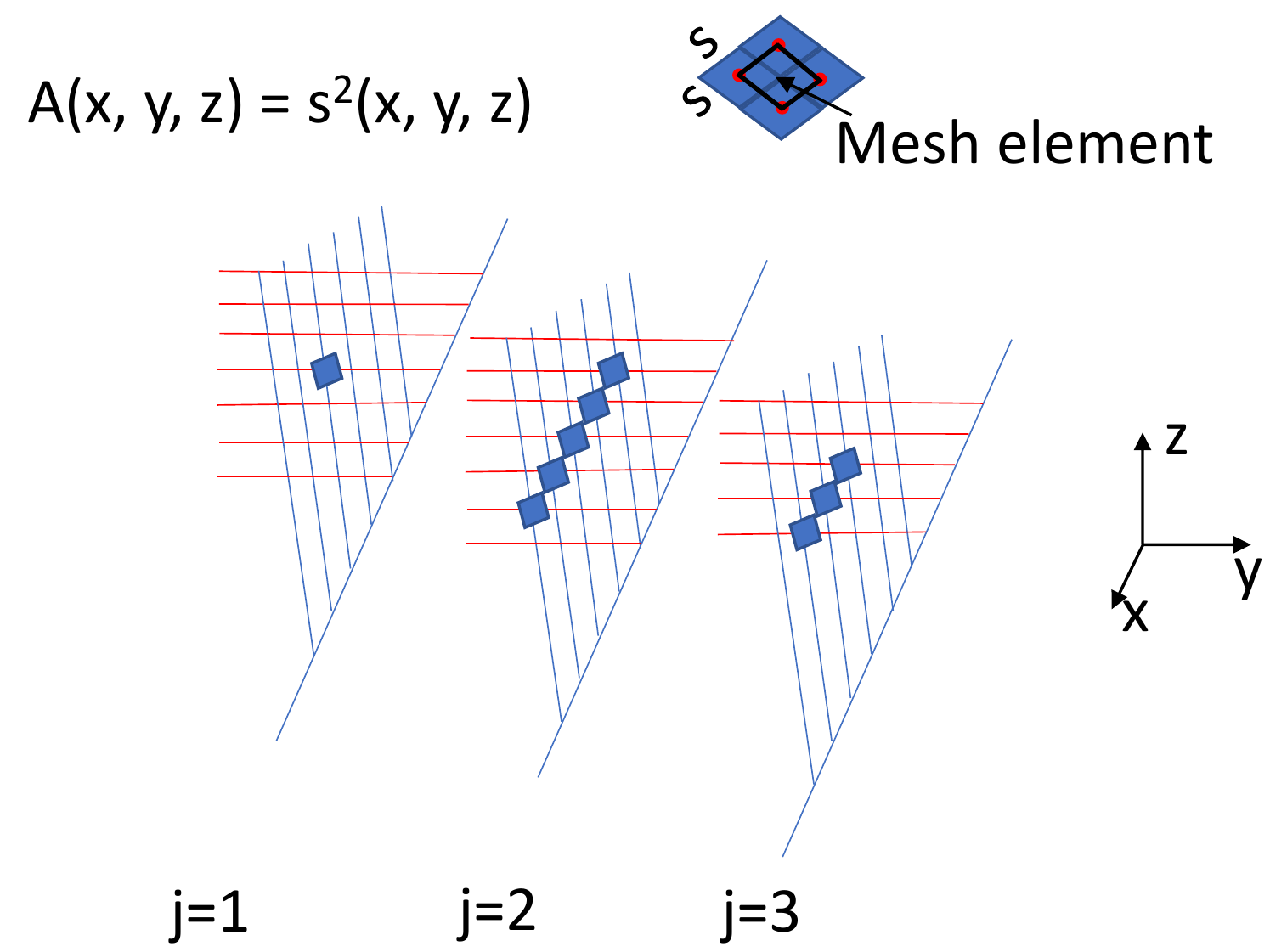}
	 \caption{}
     \end{subfigure}
      \caption{Estimated area and shape of the target based on the blocked laser beams at mesh positions. (a) Blocked laser beam intersections are shown in red at respective mesh, (b) estimated area at each blockage.} \label{Fig:detect_new}
\end{figure*}

\begin{algorithm*}[!t]
\small
\caption{Extraction of shape features of the target.}\label{Alg:features}
\begin{algorithmic}[1]
\Procedure{Shape-Features-Extraction}{}\\
\%~Consider that we are at $j^{\rm th}$ laser mesh at the $i^{\rm th}$ steering position (see Fig.~\ref{Fig:detect_new}), and each intersection of the laser beams are unique in the $(x,y,z)$ coordinate plane 
\For {$i = -N:N$}
\For {$j = 1:L$}
\If  {there is blockage at an intersection of the laser beams at $(x, y, z)$ detected using $S/n_{\rm G} < \gamma$ at respective RXs}
\State Assign area $A(x,y,z)$ to each blocked laser mesh intersection position as shown in Fig.~\ref{Fig:detect_new}(b)
\If {blockage at same $(x,z)$ position at different $j$ mesh}
\State Extend a 3D area using length $l_{\rm e}^{c}$~(where $l_{\rm e}^{c} = \Delta y$) across the $j$ mesh where the blocked positions have the same $(x,z)$. Label this 3D area as the central section as shown in Fig.~\ref{Fig:detect_new2}
\State After a central section has been identified, the area on the sides of it (i.e. different $x$ values) is labeled as wings section~(discussed in Section~\ref{Section:Features})
\If {the area of the wings decreases more than or equal to half at a later $j^{\rm th}$ mesh}
\State It is identified as the tail section. 
\State The 3D area of wings and tail section is obtained by extending the 2D area of the plane shown in Fig.~\ref{Fig:detect_new2} by $l_{\rm e}^{\rm w}$, and $l_{\rm e}^{\rm t}$, respectively, on both sides of the plane.
\State The 3D area for wings and tails is extended to later $j$ mesh, only if later mesh positions have same $(x,z)$ coordinates 
\EndIf
\EndIf
\EndIf
\EndFor
\EndFor\\
\Return{Estimated shape features of the target section-wise (if the target is detected)}
\EndProcedure
\end{algorithmic}
 \end{algorithm*}
If there is a blockage to the path of a single laser beam or a bunch of beams due to a potential target, the received intensity, $I^{(\rm RX)} (r^{(\rm RX)},d_{i,j})$ of each beam is reduced and a subsequent reduction occurs in the SNR, $S/n_{\rm G}$. This reduction is due to the reflection and absorption of the incident intensity from a target given in \ref{Eq:inten}. Let $\gamma$ be the threshold for minimum SNR. The detection threshold $\gamma$ is obtained using Neyman-Pearson decision rule and square law detection function for a given probability of false alarm~(pfa). If $S/n_{\rm G} < \gamma$, then, this is perceived as the presence of a target at that particular RX position. The material of the target does not significantly change the blockage characteristics.  

\begin{figure*}[!t]
	\centering
	\includegraphics[width=0.49\textwidth]{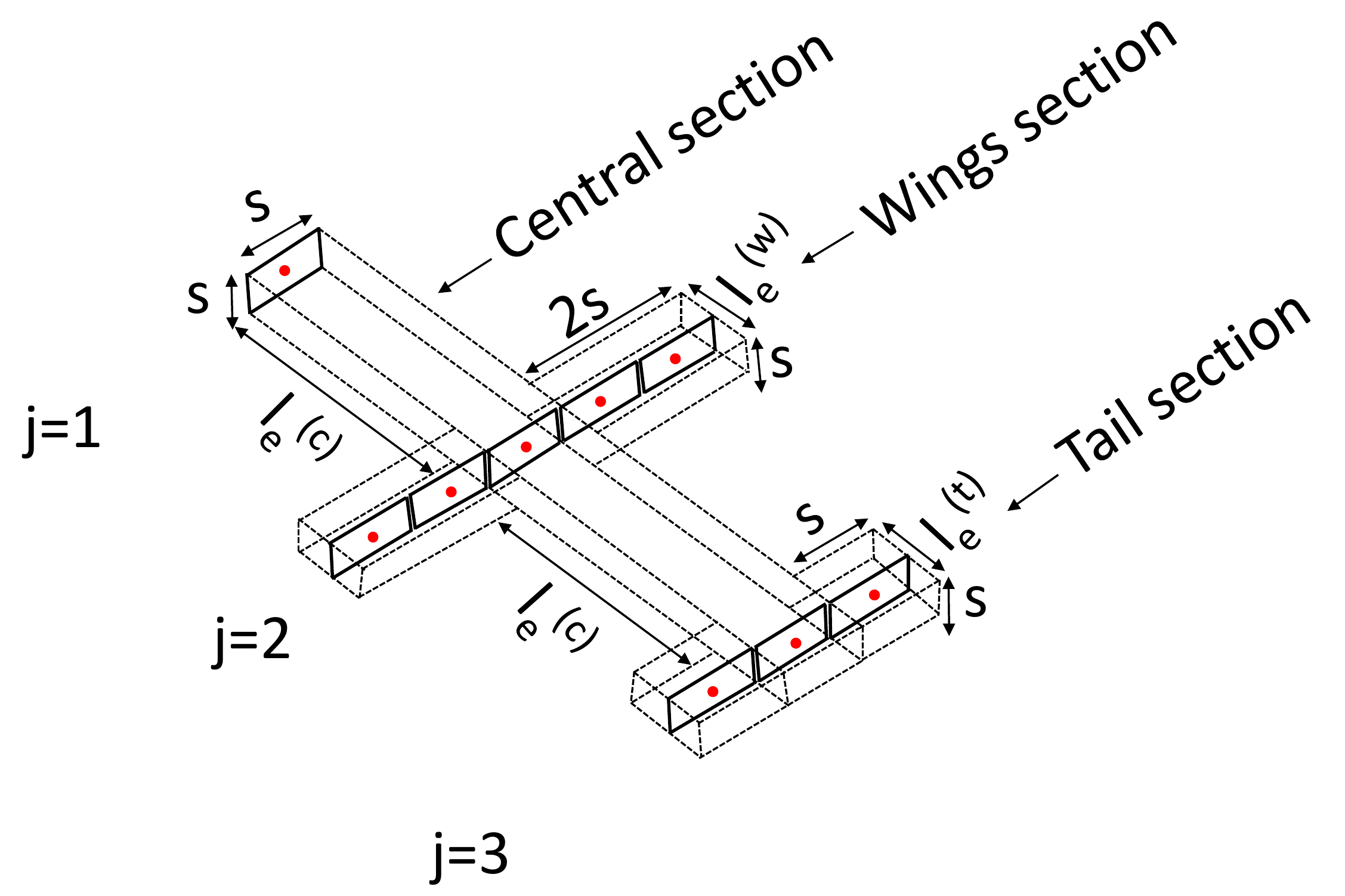} 
	\caption{The estimated 3D shape of the target based on blocked laser mesh intersections in Fig.~\ref{Fig:detect_new}.} \label{Fig:detect_new2} 
\end{figure*}

\section{Target Features} \label{Section:Features}
In this section, details of the features, i.e. 3D shape, maximum velocity, pitch, and drift angles, and maximum altitude,  obtainable by the proposed approach are provided.

\subsection{Features Associated with Target Shape} \label{Section:Shape_target}

Depending on the distance between consecutive beams, the blockage at the intersection position of laser beams on a mesh due to a target has a corresponding blocked area. Fig.~\ref{Fig:detect_new}(a) shows laser beams blocked at different positions on laser meshes, $j=1,2,3$, due to a target. The area of each blocked position~(i.e., blocked intersection of two beams) is represented as $A(x,y,z) = s^2 (x,y,z)$, where $s$ is one edge of the squares shown in Fig.~\ref{Fig:detect_new}(b). An example blockage of four neighboring intersections, their corresponding areas, and area of elements are illustrated in Fig.~\ref{Fig:detect_new}(b), where $s\approx \Delta x$ (see Fig.~\ref{Fig:Laser_mesh_new}). There is a single blocked intersection at $j=1$ laser mesh, whereas, five and three blocked intersections at $j=2$ and $j=3$, respectively. A 3D area of the target based on the blocked laser beams at $j=1,2,3$ can be approximately calculated (Fig.~\ref{Fig:detect_new2}). The overall procedure for 3D area calculation and extraction of shape features of the target are given in Algorithm~\ref{Alg:features}. 

At lines $2$, and $3$ of Algorithm~\ref{Alg:features}, we traverse over all the steering positions and laser mesh at these steering positions. We will focus on the $i^{\rm th}$ steering position and corresponding $j={1,2,3}$, laser mesh, shown in Fig.~\ref{Fig:detect_new} for the explanation of the algorithm. At lines $5$ and $6$, if there is a blockage due to the presence of a target and the condition $S/n_{\rm G} < \gamma$ is true (see Section~\ref{Section:Target_detection}), a fixed area $A(x,y,z)$ is assigned to each blocked laser mesh intersection position. The area assignment due to blockage is shown in Fig.~\ref{Fig:detect_new}(b). On line $7$ it is checked whether the blockage is present at same points in the $xz$ plane at different $j^{\rm th}$ laser mesh. To understand that, consider that we have blockages at the same points in the $xz$ plane at $j=1,2,3,$ laser mesh shown in Fig.~\ref{Fig:detect_new}(a) and (b), then we can draw a straight line passing through these points (along the $y$-axis) at $j=1,2,3,$ laser mesh and the line will be perpendicular to the laser mesh $xz$ plane. If instead of a line, we consider blockage area $A(x,y,z)$ extension across the laser mesh positions along the $y$-axis, a 3D central section is formed shown in Fig.~\ref{Fig:detect_new2}. The length of the central section between any two laser mesh is $l_{\rm e}^{c} = \Delta y$ given at line $8$. After the central section has been identified, any laser mesh intersections that are blocked and not part of the central section and extend along the $x$ or $z$-axes only is considered either as wing or tail section. The 3D area and lengths of the wings and tail sections are provided at lines $9$ to $13$ of Algorithm~\ref{Alg:features}.

\begin{figure*}[!t]
	\centering
	\includegraphics[width=0.8\textwidth]{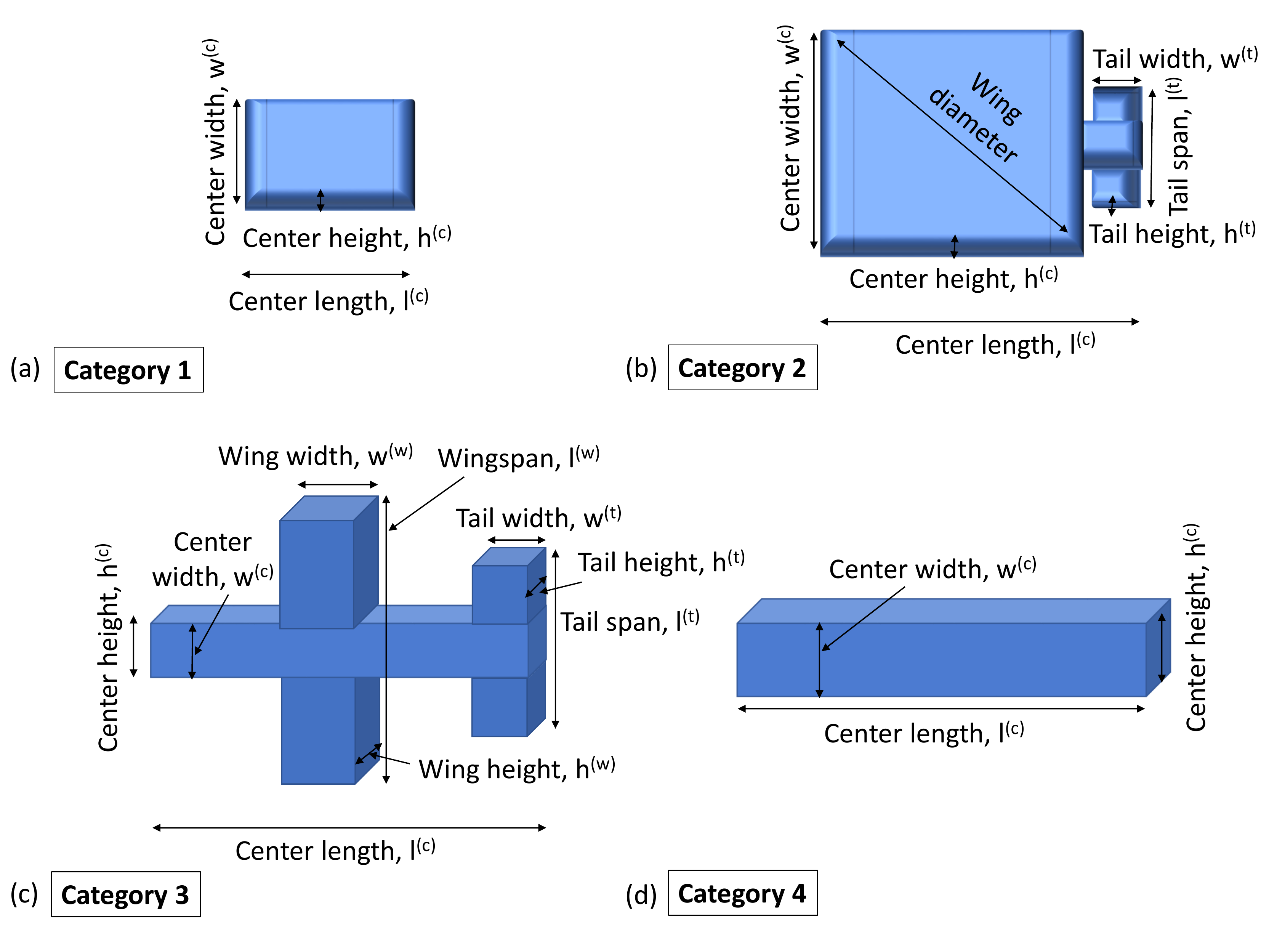} 	\caption{Four categories of the targets based on the shape.  }\label{Fig:categories} 
\end{figure*}

The 3D shape of a target can have a minimum of one section and a maximum of three sections. These are central, wing, and tail sections as shown in Fig.~\ref{Fig:detect_new2}. The total length of the central section of a target is given as $l^{(\rm c)} = l_{\rm e}^{\rm c}\times (N_{\rm B}-1)$, where $ l_{\rm e}^{\rm c}=\Delta y$, and $N_{\rm B}$ is the number of meshes that have blocked beam or beams. Four shape categories recognizable by the proposed method are shown in Fig.~\ref{Fig:categories}. The width of the central section at the $j^{\rm th}$ mesh is $w_j^{\rm (c)} = s\times N_{j,\rm B}^{(\rm c)}$, where $N_{j,\rm B}^{(\rm c)}$  are the number of the blocked laser beams at the center section of the $j^{\rm th}$ mesh. The coordinates of the blocked positions at the $j^{\rm th}$ mesh constituting the width of the central section are $\big(x_j^{(\rm c)},y_{j,\rm o}^{(\rm c)},z_{j,\rm o}^{(\rm c)}\big)$, where $y_{j,\rm o}^{(\rm c)}$, and $z_{j,\rm o}^{(\rm c)}$ are the constant coordinates in the $y$ and $z$ planes, respectively, whereas $x_j^{(\rm c)}$ varies. The height of the central section at the $j^{\rm th}$ mesh is obtained similarly as  $h_j^{(\rm c)} = s\times N_{j,\rm B}^{(\rm c)}$. The coordinates of the blocked positions at the $j^{\rm th}$ mesh contributing to the height of the central section are $\big(x_{j,\rm o}^{(\rm c)},y_{j,\rm o}^{(\rm c)},z_j^{\rm (c)}\big)$ where $z_j^{\rm (c)}$ only varies and the other two coordinate values are constant. The width and height of the central section are shown in Fig.~\ref{Fig:detect_new2} for a particular target and in Fig.~\ref{Fig:categories} for different categories of the targets. In Fig.~\ref{Fig:detect_new2}, the length, width, and height of the central section is $2 l_{\rm e}^{\rm c}$, $s$, and $s$, respectively. 

\begin{table*}
\caption{Training data of 3D shape, maximum velocity, pitch and drift angles, and maximum altitude of targets and their respective categorization based on shape.}
\centering
\resizebox{\textwidth}{!}{
\begin{tabular}{|M{2.0cm}|M{1.7cm}|M{1.3cm}|M{1.3cm}|M{1.3cm}|M{1.3cm}|M{1.3cm}|M{1.3cm}|M{1.3cm}|M{1.3cm}|M{1.3cm}|M{1.3cm}|M{1.3cm}|M{1.3cm}|}
\hline
Target Type& & Given target & Multi-rotor UAVs & Heli-copters & Fixed-wing UAVs & Small fixed-wing planes & Large fixed-wing planes & Fighter jets & Cruise missiles & Birds & Ballastic missiles & Rockets and artillery shells& HGVs  \\ \hline
Category&& cat. 3 & cat. 1& cat. 2 & cat. 3 & cat. 3 & cat. 3 & cat. 3 & cat. 3 & cat. 3 & cat. 4 & cat. 4 & cat. 4 \\ \hline
\multirow{ 3}[17]{2.5cm}{Central part} &Length (m)&$\mu = 5.65, \sigma = 0.05$& $\mu = 1.94, \sigma = 1.3$ & $\mu = 21.4, \sigma = 16.2$ & $\mu = 7.8, \sigma = 6.15$ & $\mu = 13.63, \sigma = 9.3$& $\mu = 56.8, \sigma = 14.6$ & $\mu = 20.26, \sigma = 6.2$ & $\mu = 6.48, \sigma = 2.6$ & $\mu = 0.69, \sigma = 0.4$ & $\mu = 14.1, \sigma = 11.1$ & $\mu = 3.3, \sigma = 2.5$ & $\mu = 5.3, \sigma = 4.5$ \\ \cline{2-14}
&Width (m)& $\mu = 0.52, \sigma = 0.03$ &$\mu = 2.15, \sigma = 1.6$ & $\mu = 21.4, \sigma = 16.2$ & $\mu = 0.73, \sigma = 0.5$ & $\mu = 2.17, \sigma = 1.04$ & $\mu = 4.4, \sigma = 1.32$ & $\mu = 3.36, \sigma = 1.1$ & $\mu = 0.7, \sigma = 0.2$ & NA & $\mu = 1.54, \sigma = 0.91$ & $\mu = 0.33, \sigma = 0.09$ & $\mu = 0.75, \sigma = 0.2$   \\  \cline{2-14}
& Height (m)&$\mu = 0.52, \sigma = 0.03$&$\mu = 0.6, \sigma = 0.36$ & $\mu = 5.35, \sigma = 2.26$ & $\mu = 2.1, \sigma = 1.89$ & $\mu = 3.5, \sigma = 1.28$ & $\mu = 14.6, \sigma = 5.9$& $\mu = 4.5, \sigma = 1.5$ & $\mu = 0.7, \sigma = 0.2$ & NA & $\mu = 1.5
, \sigma = 0.91$ & $\mu = 0.23, \sigma = 0.21$ & $\mu = 0.75, \sigma = 0.2$   \\  \cline{1-14}
\multirow{ 4}[27]{2.5cm}{Wings and tail sections span and width}&Wing span (m)&$\mu = 2.67, \sigma = 0.04$ &NA& $\mu = 21.35, \sigma = 13.99$ & $\mu = 20.7, \sigma = 17.36$ & $\mu = 15.2, \sigma = 7.9$ & $\mu = 59.4, \sigma = 17.4$ & $\mu = 12, \sigma = 4.44$ & $\mu = 2.5, \sigma = 0.52$& $\mu = 1.4, \sigma = 0.7$ & NA & NA & NA  \\ \cline{2-14}
&Wing width (m) & $\mu = 0.75, \sigma = 0.02$ &NA & NA & $\mu = 1.03, \sigma = 0.62$ & $\mu = 1.93, \sigma = 0.94$ & $\mu = 7.15, \sigma = 2.36$ & $\mu = 3.7, \sigma = 1$ & $\mu = 0.8, \sigma = 0.4$& $\mu = 0.8, \sigma = 0.5$ & NA & NA& NA   \\  \cline{2-14}
& Tail span (m) &$\mu = 0.89, \sigma = 0.035$& NA & $\mu = 3.1, \sigma = 1.2$ & $\mu = 6.56, \sigma = 5.5$ & $\mu = 4.7, \sigma = 2.8$ & $\mu = 22, \sigma = 8$ & $\mu = 4.6, \sigma = 2.02$ & $\mu = 0.9, \sigma = 0.2$& $\mu = 0.41, \sigma = 0.2$ & NA& NA& NA   \\  \cline{2-14}
& Tail width (m) &$\mu = 0.27, \sigma = 0.022$& NA & $\mu = 1, \sigma = 0.7$ & $\mu = 1.85, \sigma = 1.6$ & $\mu = 1.9, \sigma = 2.2$ & $\mu = 8.33, \sigma = 5.5$ &$\mu = 1.87, \sigma = 0.9$ & $\mu = 0.3, \sigma = 0.15$ & $\mu = 0.18, \sigma = 0.12$ & NA & NA & NA   \\ \cline{1-14}
$v^{(\rm max)}$ (m/s)&&$\mu = 350.2, \sigma = 7.5$&$\mu = 20.2, \sigma = 8.5$ & $\mu = 65.9, \sigma = 28.14$ & $\mu = 76.8, \sigma = 62.3$ & $\mu = 154.5, \sigma = 87$ & $\mu = 266.5, \sigma = 17.8$&  $\mu = 636, \sigma = 99.2$ & $\mu = 800, \sigma = 750$ & $\mu = 32.2, \sigma = 29.5$ & $\mu = 3500, \sigma = 3000$ & $\mu = 937.5, \sigma = 593$ & $\mu = 4000, \sigma = 2500$ \\ \cline{1-14}
$\alpha^{\circ}$& &$\mu = 12, \sigma = 8$ & $\mu = 10, \sigma = 5$ & $\mu = 15, \sigma = 25$ & $\mu = 20, \sigma = 30$ & $\mu = 10, \sigma = 20$ & $\mu = 8, \sigma = 15$ & $\mu = 20, \sigma = 45$ & $\mu = 12, \sigma = 30$ & $\mu = 10, \sigma = 30$ & $\mu = 5, \sigma = 30$ & $\mu = 5, \sigma = 15$ & $\mu = 10, \sigma = 60$ \\ \cline{1-14}
$\beta^{\circ}$ &&$\mu = 7, \sigma = 5$& $\mu = 15, \sigma = 10$&$\mu = 10, \sigma = 15$ & $\mu = 15, \sigma = 20$ & $\mu = 8, \sigma = 15$ & $\mu = 7, \sigma = 15$ & $\mu = 15, \sigma = 40$ & $\mu = 8, \sigma = 10$ & $\mu = 5, \sigma = 15$ & $\mu = 5, \sigma = 20$ & $\mu = 5, \sigma = 8$ & $\mu = 8, \sigma = 45$ \\ \cline{1-14}
$h^{\rm G}$ (km) & & $\mu = 30, \sigma = 25$&$\mu = 4.62, \sigma = 2.21$ & $\mu = 4.7, \sigma = 1.9$ & $\mu = 11.5, \sigma = 6.9$ & $\mu = 9.95, \sigma = 5.31$ & $\mu = 14, \sigma = 1.15$ & $\mu = 17.2, \sigma = 1.3$ & $\mu = 20, \sigma = 17.5$ & $\mu = 2.8, \sigma = 2$& $\mu = 721, \sigma = 829$ & $\mu = 13.2, \sigma = 14.63$ & $\mu = 23, \sigma = 10.5$ \\ \hline
\end{tabular} \label{Table:Train_Data}
}
\end{table*}

The wingspan at a $j^{\rm th}$ mesh from Fig.~\ref{Fig:detect_new2} and Fig.~\ref{Fig:categories} is given as $l^{\rm (w)} = s\times N_{j,\rm B}^{(\rm w)}$, where $N_{j,\rm B}^{(\rm w)}$ are the number of blocked laser beam positions of the $j^{\rm th}$ mesh at the wings section. The coordinates of the blocked positions at  the $j^{\rm th}$ mesh for the wingspan are $\big(x_j^{(\rm w)},y_{j,\rm o}^{(\rm w)},z_{j,\rm o}^{(\rm w)}\big)$. For example in Fig.~\ref{Fig:detect_new2}, the wingspan is $5s$. The width of the wings section at the $j^{\rm th}$ mesh is $w^{(\rm w)} = l_{\rm e}^{\rm (w)}\times N_{j,\rm B}^{(\rm w)}$, where $ l_{\rm e}^{\rm (w)}$ is a constant value, and for simplicity can be taken as $ l_{\rm e}^{\rm (w)}=s$, and coordinates of the blocked positions corresponding to the width of the wings at the $j^{\rm th}$ mesh are $\big(x_{j,\rm o}^{(\rm w)},y_{j}^{(\rm w)},z_{j,\rm o}^{\rm (w)}\big)$. The height of the wings section at the $j^{\rm th}$ mesh is given as $h^{(\rm w)} = sN_{j,\rm B}^{(\rm w)}$. The coordinates of the blocked mesh positions forming the height of the wings are represented as $\big(x_{j,\rm o}^{(\rm w)},y_{j,\rm o}^{(\rm w)},z_j^{\rm (w)}\big)$.  The width and height of the wings section for a given target is shown in Fig.~\ref{Fig:detect_new2}, and for Category~3 in Fig.~\ref{Fig:categories}. In Fig.~\ref{Fig:detect_new2}, the width and height of the wings section is $ l_{\rm e}^{\rm (w)}$, and $s$, respectively. The dimensions of the tail section are obtained similarly as for the wings section.

\subsection{Velocity, Pitch and Drift Angles, and Altitude of the Target} \label{Section:local_map}
The motion characteristics of a target at a given time depends on the velocity, and pitch and drift angles. These three features of a target can be estimated using the proposed framework and can be utilized to classify a target. Let $\Delta d_{i,i+1}$ represents the distance between any two steering positions, $i$ and $i+1$, which have one or more blocked intersections due to a target, and $\Delta t_{i,i+1}$ is the corresponding time difference for a target to move between mesh locations at $i^{\rm th}$ and $(i+1)^{\rm th}$ steering positions. The instantaneous velocity is represented as $v(t) = \frac{\Delta d_{i,i+1}}{\Delta t_{i,i+1}}$. Over the steering positions, we can write the maximum velocity as 
\begin{equation}
 v^{(\rm max)} = \max \bigg(\frac{ \Delta d_{i,i+1}}{\Delta t_{i,i+1}}\bigg),~\forall~ i = -N, -N+1,\hdots, N-1, N.
\end{equation}

\begin{table*}
\caption{Size, velocity, and flying altitude of popular rotary and fixed-wing UAVs.}
  \centering
\resizebox{\textwidth}{!}{

\begin{tabular}{|p{0.005cm}|p{0.005cm}|p{0.005cm}|p{0.005cm}|p{0.005cm}|p{0.005cm}|p{0.005cm}|}
\hline
\multicolumn{1}{|c|}{Target Type}&\multicolumn{1}{|c|}{Type}&\multicolumn{3}{|c|}{Total area}&\multicolumn{1}{|c|}{ $v^{(\rm max)}$ (m/s)}&\multicolumn{1}{|c|}{Max. altitude (km)} \\
			\hline
        \multicolumn{1}{|c|}{}& \multicolumn{1}{|c|}{}&\multicolumn{1}{|c|}{Length of central part (m)}&\multicolumn{1}{|c|}{Wing span (m)}&\multicolumn{1}{|c|}{Height of central part (m)}&\multicolumn{1}{|c|}{}&\multicolumn{1}{|c|}{}\\
             \hline
              \multicolumn{1}{|c|}{Phantom 4~\cite{Phantom4}}& \multicolumn{1}{|c|}{Multi-rotor UAV}&\multicolumn{1}{|c|}{$0.4$}&\multicolumn{1}{|c|}{$0.15$, propellers dia}&\multicolumn{1}{|c|}{$0.19$}&\multicolumn{1}{|c|}{$20$}&\multicolumn{1}{|c|}{$0.5$}\\
             \hline
 	\multicolumn{1}{|c|}{Matrice 600~\cite{Matrice600}}& \multicolumn{1}{|c|}{Multi-rotor UAV}&\multicolumn{1}{|c|}{$1.18$}&\multicolumn{1}{|c|}{$0.5$, propellers dia}&\multicolumn{1}{|c|}{$0.5$}&\multicolumn{1}{|c|}{$18$}&\multicolumn{1}{|c|}{$2.5$}\\
             \hline
\multicolumn{1}{|c|}{Amazon Prime Air~\cite{Amazon_UAV}}& \multicolumn{1}{|c|}{Multi-rotor UAV}&\multicolumn{1}{|c|}{$9.1$}&\multicolumn{1}{|c|}{$2.5$, frame and propellers}&\multicolumn{1}{|c|}{$1.5$}&\multicolumn{1}{|c|}{$22.2$}&\multicolumn{1}{|c|}{$0.15$}\\
             \hline
\multicolumn{1}{|c|}{MQ-8B~\cite{MQ8}]}& \multicolumn{1}{|c|}{Helicopter UAV}&\multicolumn{1}{|c|}{$7.3$}&\multicolumn{1}{|c|}{$8.4$, propellers dia}&\multicolumn{1}{|c|}{$2.9$}&\multicolumn{1}{|c|}{$59.2$}&\multicolumn{1}{|c|}{$6.1$}\\
             \hline
\multicolumn{1}{|c|}{Sky Surfer~\cite{Skysurfer}}& \multicolumn{1}{|c|}{Fixed-wing UAV}&\multicolumn{1}{|c|}{$0.91$}&\multicolumn{1}{|c|}{$1.4$}&\multicolumn{1}{|c|}{$0.32$}&\multicolumn{1}{|c|}{$13.8$}&\multicolumn{1}{|c|}{$0.2$}\\
             \hline
\multicolumn{1}{|c|}{Eclipse 2.0~\cite{FW1}}& \multicolumn{1}{|c|}{Fixed-wing UAV}&\multicolumn{1}{|c|}{$1.1$}&\multicolumn{1}{|c|}{$5.5$}&\multicolumn{1}{|c|}{$0.25$}&\multicolumn{1}{|c|}{$28$}&\multicolumn{1}{|c|}{$0.13$}\\
             \hline
\multicolumn{1}{|c|}{MQ-9A~\cite{MQ9}}& \multicolumn{1}{|c|}{Fixed-wing UAV}&\multicolumn{1}{|c|}{$11$}&\multicolumn{1}{|c|}{$20$}&\multicolumn{1}{|c|}{$3.6$}&\multicolumn{1}{|c|}{$123.5$}&\multicolumn{1}{|c|}{$15.24$}\\
             \hline
\multicolumn{1}{|c|}{Global Hawk~\cite{GlobalHawk}}& \multicolumn{1}{|c|}{Fixed-wing UAV}&\multicolumn{1}{|c|}{$14.5$}&\multicolumn{1}{|c|}{$35.4$}&\multicolumn{1}{|c|}{$4.7$}&\multicolumn{1}{|c|}{$175$}&\multicolumn{1}{|c|}{$19.8$}\\
             \hline
                \end{tabular}
                
\label{Table:Train_Data2}
}
\end{table*}

The trajectory variations of a target in the elevation and azimuth planes can be represented using pitch and drift angles, respectively. Let $\alpha (t)$, and $\beta (t)$ represent the pitch and drift angles of a target, respectively. If $h_{i+1}(t)$ and $h_{i}(t)$ are the estimated heights of the target at $i+1$ and $i$ steering positions, $x_{i+1}(t)$ and $x_i(t)$ are the $x$-coordinates of the target at $i+1$ and $i$ steering positions at time $t$, and $\Delta P$ is the distance between two consecutive steering positions shown in Fig.~\ref{Fig:Laser_steer}, then $\alpha (t)$ and $\beta (t)$ are given as~\cite{laser44,laser55} 
\begin{align}
&\alpha (t) = \tan^{-1}{\Bigg( \frac{h_{i+1}(t) - h_{i}(t)}{\Delta P }}\Bigg), \\
&\beta (t) = \tan^{-1}{\Bigg( \frac{x_{i+1}(t) - x_{i}(t)}{\Delta P}}\Bigg).
\end{align}
The maximum altitude of an aerial target contains valuable information characterizing the target. This feature can also be determined using our proposed setup. The maximum z-coordinate value at the intersection pair of blocked beams provides the maximum altitude of the target at that location. The maximum altitude of the target from the ground is represented as $h^{(\rm G)}$. The maximum altitude, velocity, pitch and drift angles, and 3D shape features for different types of targets are given in Table~\ref{Table:Train_Data}. The data in Table~\ref{Table:Train_Data} is used for creating the training set that is utilized developing the ML models for classification. Another Table~\ref{Table:Train_Data2} is provided that contains the size, maximum velocity and flight altitude of popular rotary and fixed-wing UAVs.

\section{Classification, Localization, and Target Tracking} \label{Section:Classification}
In this section, details of  the classification, localization, and target tracking models used in our approach are provided.

\subsection{Classification of a Target using Training Data}

\begin{figure*}[!t] 
    \centering
	\begin{subfigure}{0.7\columnwidth}
    \centering
	\includegraphics[width=\textwidth]{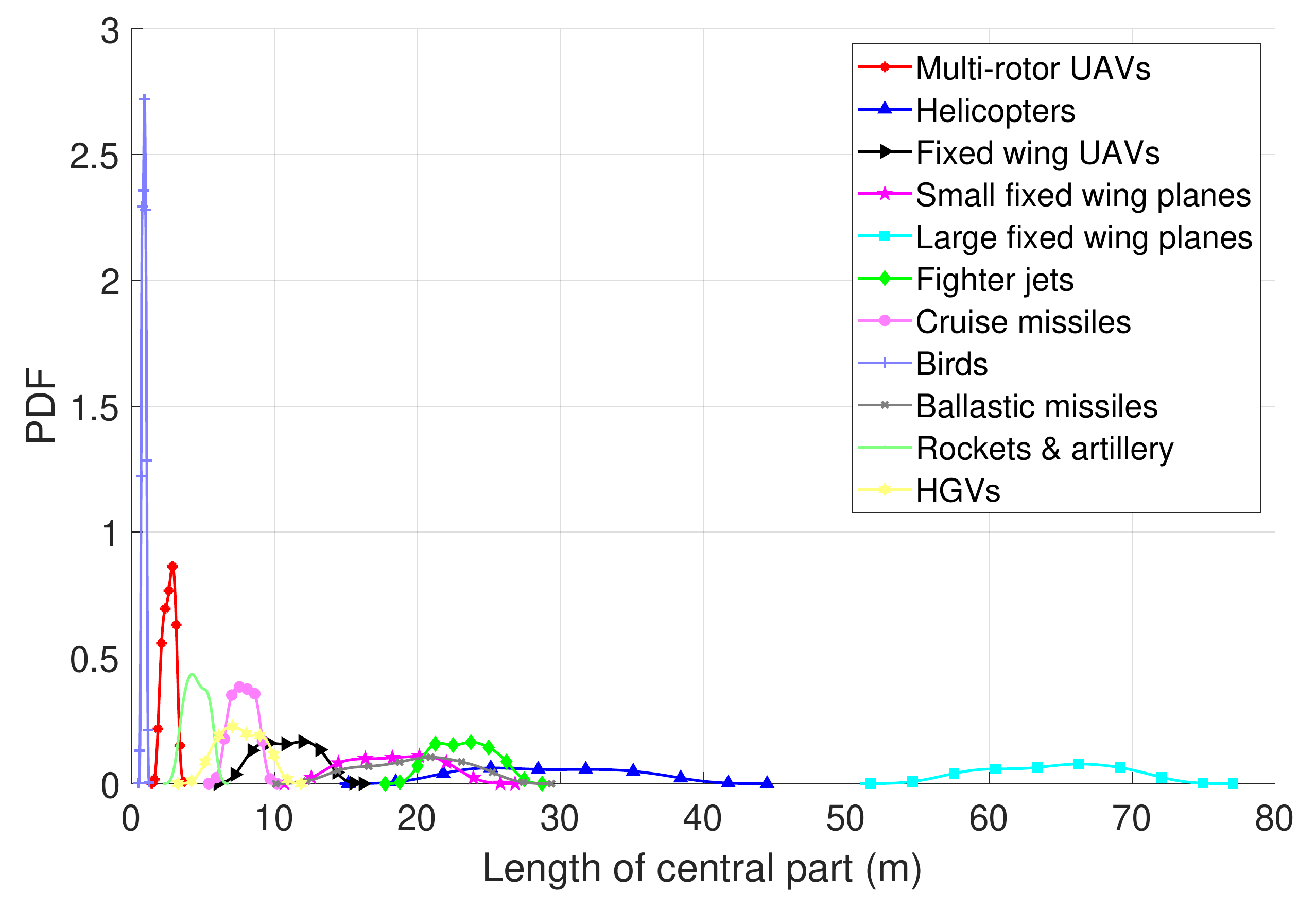}
	  \caption{}  
    \end{subfigure}    
    \begin{subfigure}{0.7\columnwidth}
    \centering
	\includegraphics[width=\textwidth]{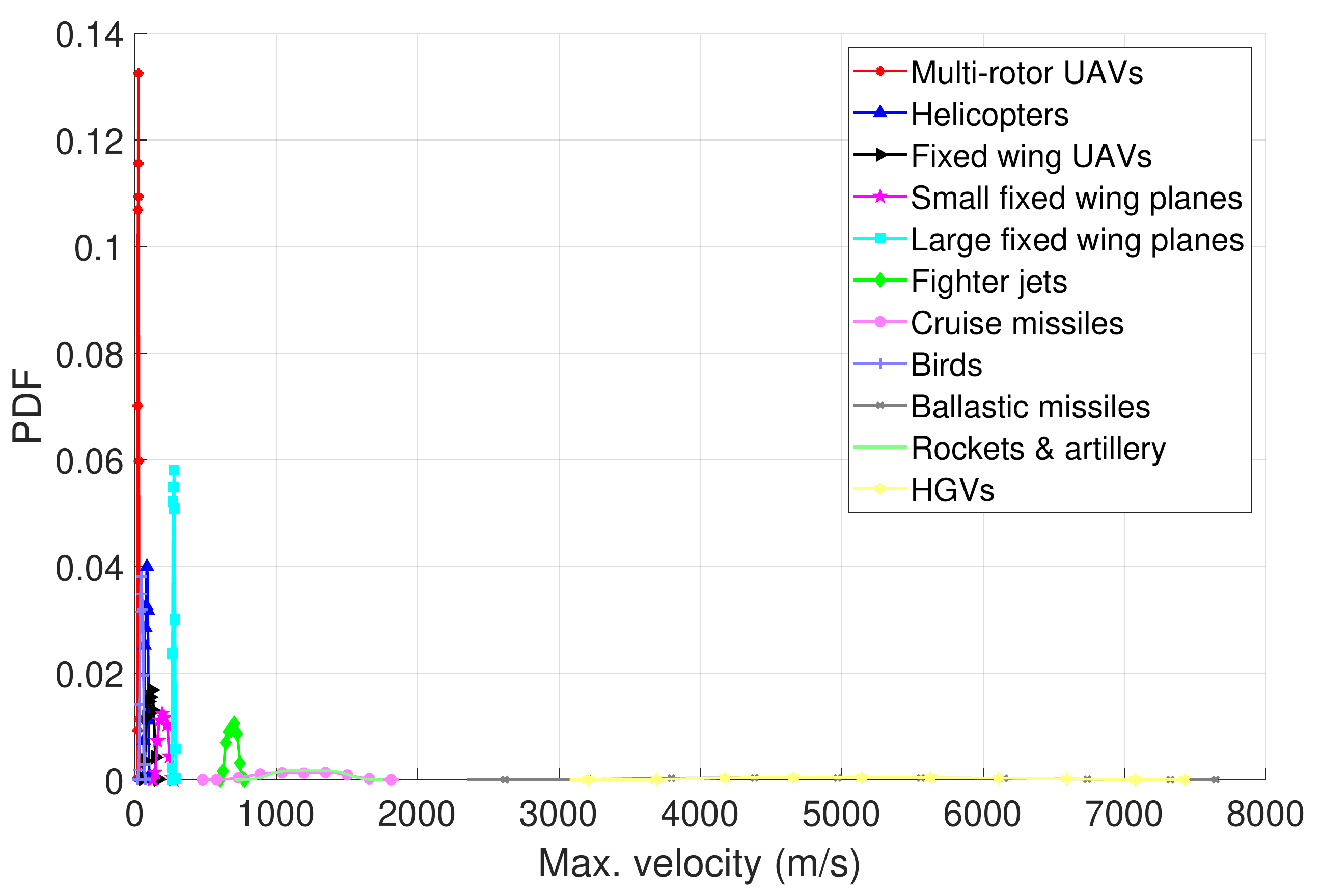}
	  \caption{}  
    \end{subfigure}
    \caption{PDFs of length of central section and maximum velocity of the training targets.} \label{Fig:PDF_len_vel}
\end{figure*}

Any flying object that disrupts the path of the laser beams is considered as a potential target in the proposed approach. These objects can be man-made or birds. Flying objects are classified based on the features discussed in Section~\ref{Section:Features}, i.e., shape, maximum velocity, pitch and drift angle characteristics, and maximum flight altitude by developing ML models using training data details provided in Table~\ref{Table:Train_Data}. Target types are grouped into four categories based on the 3D shape~(Fig.~\ref{Fig:categories}), i.e., drone-like objects, chopper-like objects, fixed-wing type objects, and missile-like objects. The dimensions of the 3D shapes are provided in Table~\ref{Table:Train_Data}. The first category is assigned to multi-rotor UAVs that have either a square or rectangular central section from the mainframe and mounted rotors onboard the mainframe. The 3D shape of the first category is shown in Fig.~\ref{Fig:categories}(a). The second category for helicopters with a square central section due to the rotor blades followed by a tail section shown in Fig.~\ref{Fig:categories}(b). The tail section is not present in some cases for Category~2. Central sections of objects in Category 2 are expected to be larger than central sections of objects in Category~1. In Fig.~\ref{Fig:categories}(c), a third category is shown that covers all the aerial targets with central, wings, and tail sections. These include fixed-wing UAVs and planes, cruise missiles, and birds. The fourth category represents targets with only a long main central section without any significant wings and tail spans~(Fig.~\ref{Fig:categories}(d)). The fourth category includes ballistic missiles, rockets and artillery shells, and  hypersonic glide vehicles~(HGVs).

The parameters of the training data in Table~\ref{Table:Train_Data} consist of the length, width, and height of the central section of eleven different targets~(i.e., classes), grouped into four categories. The wingspan, wing width, tail span, and tail width values are also provided. The height of the wings and tail sections have small variances among different types of targets, hence, these two features are not included in the training set to decrease the statistical noise as low as possible. Other features used while creating the training set are maximum velocity, pitch and drift angles, and maximum altitude values of eleven different types. Each and every feature that are used creating the training data assumed to have a Gaussian distribution with the mean and standard deviations provided in Table~\ref{Table:Train_Data}. The \textit{NA} entries in Table~\ref{Table:Train_Data} indicate that there are no values present for the training features of that particular target type. 

\begin{figure*}[!t]
	\centering
	\includegraphics[width=0.75\columnwidth]{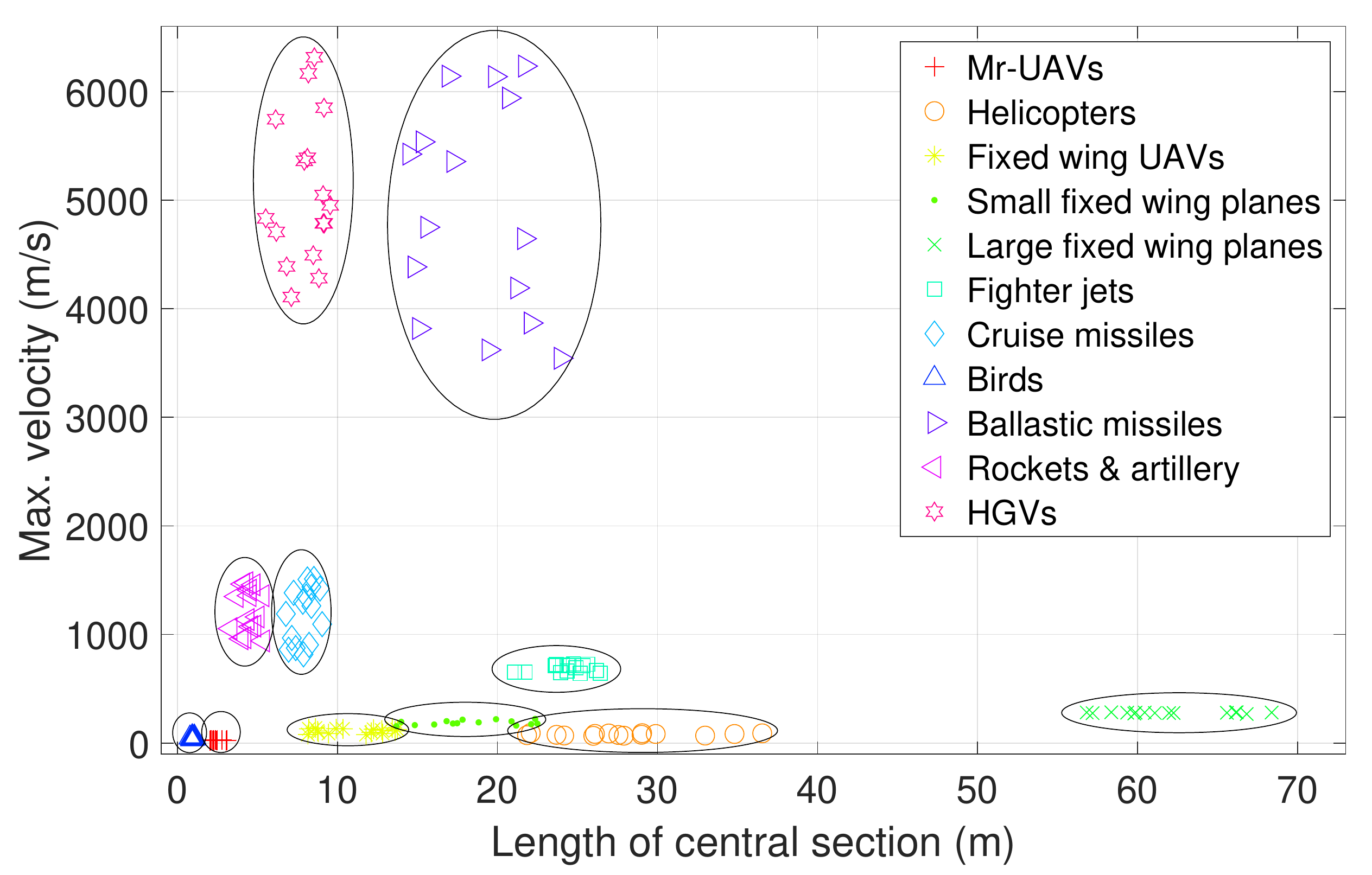} 
	\caption{Visualization of the length of the central section and maximum velocity of training targets. } \label{Fig:visual_boundry} 
\end{figure*}

Fig.~\ref{Fig:PDF_len_vel} shows the probability density functions~(PDFs) of the length of the central section and maximum velocity of different types of training targets based on the parameters provided in Table~\ref{Table:Train_Data}. Even though the PDFs of different training targets overlaps, the used ML models take several other features into account that yields to high classification accuracy. Visual boundaries among the different types of targets for length of the central section and maximum velocity are given in Fig.~\ref{Fig:visual_boundry}. It is observed that even with only two features, target classes become quite separable. Let $\textbf{M}^{(\rm Tr)}$ represent the matrix containing training data of eleven targets that is given as
\begin{equation} \label{Eq:Mat_train}
\textbf{M}^{(\rm Tr)} = 
\!\begin{aligned}
&
\left[\begin{matrix}
         l_1^{(\rm c, C_1)}&  w_1^{(\rm c, C_1)} &  h_1^{(\rm c, C_1)} &  l_1^{(\rm w, C_1)} &  w_1^{(\rm w, C_1)} & l_1^{(\rm t, C_1)}\\
	l_2^{(\rm c, C_1)}&w_2^{(\rm c, C_1)}&h_2^{(\rm c, C_1)}&l_2^{(\rm w, C_1)}&w_2^{(\rm w, C_1)}&l_2^{(\rm t, C_1)}\\
	\vdots & \vdots &\vdots &\cdots& \vdots& \vdots\\
	l_k^{(\rm c, C_1)}&w_k^{(\rm c, C_1)}&h_k^{(\rm c, C_1)}&l_k^{(\rm w, C_1)}&w_k^{(\rm w, C_1)}&l_k^{(\rm t, C_1)}\\
	l_1^{(\rm c, C_2)}&w_1^{(\rm c, C_2)}&h_1^{(\rm c, C_2)}&l_1^{(\rm w, C_2)}&w_1^{(\rm w, C_2)}&l_1^{(\rm t, C_2)}\\
	l_2^{(\rm c, C_2)}&w_2^{(\rm c, C_2)}&h_2^{(\rm c, C_2)}&l_2^{(\rm w, C_2)}&w_2^{(\rm w, C_2)}&l_2^{(\rm t, C_2)}\\
	\vdots & \vdots &\vdots &\cdots& \vdots& \vdots\\
	l_k^{(\rm c, C_2)}&w_k^{(\rm c, C_2)}&h_k^{(\rm c, C_2)}&l_k^{(\rm w, C_2)}&w_k^{(\rm w, C_2)}&l_k^{(\rm t, C_2)}\\
	\vdots & \vdots &\vdots &\cdots& \vdots& \vdots\\
	\vdots & \vdots &\vdots &\cdots& \vdots& \vdots\\
	l_1^{(\rm c, C_K)}&w_1^{(\rm c, C_K)}&h_1^{(\rm c, C_K)}&l_1^{(\rm w, C_K)}&w_1^{(\rm w, C_K)}&l_1^{(\rm t, C_K)}\\
	l_2^{(\rm c, C_K)}&w_2^{(\rm c, C_K)}&h_2^{(\rm c, C_K)}&l_2^{(\rm w, C_K)}&w_2^{(\rm w, C_K)}&l_2^{(\rm t, C_K)}\\
	\vdots & \vdots &\vdots &\cdots& \vdots& \vdots\\
	l_k^{(\rm c, C_K)}&w_k^{(\rm c, C_K)}&h_k^{(\rm c, C_K)}&l_k^{(\rm w, C_K)}&w_k^{(\rm w, C_K)}&l_k^{(\rm t, C_K)}
\end{matrix}\right.\\
&\qquad\qquad
\left.\begin{matrix}
  w_1^{(\rm t, C_1)}& v_1^{(\rm C_1)}& \alpha_1^{(\rm C_1)}& \beta_1^{(\rm C_1)}& h_1^{(\rm G, C_1)}\\
w_2^{(\rm t, C_1)}&v_2^{(\rm C_1)}&\alpha_2^{(\rm C_1)}&\beta_2^{(\rm C_1)}&h_2^{(\rm G, C_1)}\\
 \vdots& \vdots& \vdots& \vdots& \vdots\\
w_k^{(\rm t, C_1)}&v_k^{(\rm C_1)}&\alpha_k^{(\rm C_1)}&\beta_k^{(\rm C_1)}&h_k^{(\rm G, C_1)}\\
w_1^{(\rm t, C_2)}&v_1^{(\rm C_2)}&\alpha_1^{(\rm C_2)}&\beta_1^{(\rm C_2)}&h_1^{(\rm G, C_2)}\\
w_2^{(\rm t, C_2)}&v_2^{(\rm C_2)}&\alpha_2^{(\rm C_2)}&\beta_2^{(\rm C_2)}&h_2^{(\rm G, C_2)}\\
 \vdots& \vdots& \vdots& \vdots& \vdots\\
w_k^{(\rm t, C_2)}&v_k^{(\rm C_2)}&\alpha_k^{(\rm C_2)}&\beta_k^{(\rm C_2)}&h_k^{(\rm G, C_2)}\\
 \vdots& \vdots& \vdots& \vdots& \vdots\\
 \vdots& \vdots& \vdots& \vdots& \vdots\\
w_1^{(\rm t, C_K)}&v_1^{(\rm C_K)}&\alpha_1^{(\rm C_K)}&\beta_1^{(\rm C_K)}&h_1^{(\rm G, C_K)}\\
w_2^{(\rm t, C_K)}&v_2^{(\rm C_K)}&\alpha_2^{(\rm C_K)}&\beta_2^{(\rm C_K)}&h_2^{(\rm G, C_K)}\\
 \vdots& \vdots& \vdots& \vdots& \vdots\\
w_k^{(\rm t, C_K)}&v_k^{(\rm C_K)}&\alpha_k^{(\rm C_K)}&\beta_k^{(\rm C_K)}&h_k^{(\rm G, C_K)}  
\end{matrix}  \right]  
 \end{aligned}  
  \end{equation} 
where $l_1^{(\rm c, C_1)}, w_1^{(\rm c, C_1)}$, and $h_1^{(\rm c, C_1)}$
are the first points representing length, width, and height of the central section, respectively, belonging to the first class. $l_1^{(\rm w, C_1)}, w_1^{(\rm w, C_1)}$ are first points of the wingspan and wing width of the first class, respectively, and the first points of the tail span and width are represented with $l_1^{(\rm t, C_1)}, w_1^{(\rm t, C_1)}$. Moreover, the first points of maximum velocity, pitch and drift angles, and maximum flight altitude are represented, respectively, with $v_1^{(\rm C_1)}, \alpha_1^{(\rm C_1)}, \beta_1^{(\rm C_1)}, h_1^{(\rm G, C_1)}$, for the first class. Each class has $k$ number of samples and there are $K=11$ classes in our training set. 

Similar to the training data, the feature values for a given target is given in the form of a matrix, $\textbf{M}^{(\rm eval)}$ as  
\begin{equation} \label{Eq:Mat_target}
\textbf{M}^{(\rm eval)} = 
\!\begin{aligned}
&
\left[\begin{matrix}
  l_1^{(\rm c)}&w_1^{(\rm c)}&h_1^{(\rm c)}&l_1^{(\rm w)}&w_1^{(\rm w)}&l_1^{(\rm t)}\\
 l_2^{(\rm c)}&w_2^{(\rm c)}&h_2^{(\rm c)}&l_2^{(\rm w)}&w_2^{(\rm w)}&l_2^{(\rm t)}\\
\vdots & \vdots &\vdots &\cdots& \vdots& \vdots\\
l_{k'}^{(\rm c)}&w_{k'}^{(\rm c)}&h_{k'}^{(\rm c)}&l_{k'}^{(\rm w)}&w_{k'}^{(\rm w)}&l_{k'}^{(\rm t)}
\end{matrix}\right.\\
&\qquad\qquad
\left.\begin{matrix}
  w_1^{(\rm t)}&v_1&\alpha_1&\beta_1&h_1^{(\rm G)}\\
w_2^{(\rm t)}&v_2&\alpha_2&\beta_2&h_2^{(\rm G)}\\
 \vdots& \vdots& \vdots& \vdots& \vdots\\
w_{k'}^{(\rm t)}&v_{k'}&\alpha_{k'}&\beta_{k'}&h_{k'}^{(\rm G)}
\end{matrix}\right]
\end{aligned}
\end{equation} 
where the number of samples for the target are $k'$. The target data is interpolated to adjust the size of the target's data equal to the training data that can be formulated as 
\begin{align}
\textbf{M}^{(\rm eval)} =& {\rm interpolate}\bigg(\textbf{M}^{(\rm eval)}, k\bigg).
\end{align}
 Let $\textbf{C}^{(\rm Tr)}$ represents an array that contains the target classes. In \ref{Eq:Mat_class}, the size of this vector corresponding to \ref{Eq:Mat_train} are $k\times K$, given as
\begin{equation} \label{Eq:Mat_class}
\textbf{C}^{(\rm Tr)} = 
\!\begin{aligned}
&
\left[\begin{matrix}
 C_{1,1}~~
C_{1,2}~~
\hdots
C_{1,k}~~
C_{2,1}~~
C_{2,2}~~
\hdots
 \end{matrix}\right.\\
&\qquad\qquad
\left.\begin{matrix}
  C_{2,k}~~
\hdots \hdots~ 
C_{K,1}~~
C_{K,2}~~
\hdots
C_{K,k}
\end{matrix}\right]^T
\end{aligned}
\end{equation} 
where $T$ is the transpose operation. 

\begin{algorithm*}[!t]
\small
\caption{Simultaneous Detection, Classification, Localization, and Tracking of the Aerial Target.}\label{Alg:SDCLM}
\begin{algorithmic}[1]
\Procedure{SDCLT}{}\\
\%~The total coverage of a single virtual door is $M\Delta x h$ and the total area spanned at all the steering positions is $M\Delta x \Delta y (2N+1) L h$\\
\%~Considering that we have $j^{}$ laser mesh at the $i^{\rm th}$ steering position (see Fig.~\ref{Fig:Laser_steer} )
\For {i = -N:N}
\If  {there are laser beams with $S/n_{\rm G} < \gamma$} for a given pfa
\State Obtain the estimated features of the target given in Table~\ref{Table:Train_Data}, (and discussed in Section~\ref{Section:Features}) 
\State Categories and classify the target (see Section~\ref{Section:Classification})
\If {a positive threat identified}
\State Obtain the localization and tracking of the target and update at every $i^{\rm th}$ steering position (see Section~\ref{Section:local_track})
\EndIf
\Else
\State Steering positions only
\EndIf
\EndFor\\
\Return{Estimated coordinates, and features of the target (if detected)}
\EndProcedure
\end{algorithmic}
 \end{algorithm*}

In this study, four different types of classifiers, i.e., Naive Bayes~(NB), Linear Discriminant Analysis~(LDA), K-nearest Neighbor~(KNN), and Random Forest~(RF), are used for classification. With $f$ representing the modeling function of a classifier, $\mathcal{M}_1$, $\mathcal{M}_2$, $\mathcal{M}_3$, and $\mathcal{M}_4$ expresses the corresponding models as given below
\begin{align}
\mathcal{M}_1 = f^{(\rm NB)}\bigg(\textbf{M}^{(\rm Tr)}, \textbf{C}^{(\rm Tr)}\bigg),\\
\mathcal{M}_2 = f^{(\rm LDA)}\bigg(\textbf{M}^{(\rm Tr)}, \textbf{C}^{(\rm Tr)}\bigg),\\
\mathcal{M}_3 = f^{(\rm KNN)}\bigg(\textbf{M}^{(\rm Tr)}, \textbf{C}^{(\rm Tr)}\bigg),\\
\mathcal{M}_4 = f^{(\rm RF)}\bigg(\textbf{M}^{(\rm Tr)}, \textbf{C}^{(\rm Tr)}\bigg).
\end{align}
The interpolated target's data and the classifier models are used to estimate the particular class of the target using the prediction function given as
\begin{align}
C^{(\rm {est}, \mathcal{M}_q)} =& {\rm predict}\bigg(\mathcal{M}_q,\textbf{M}^{(\rm eval)}\bigg),
\end{align}
where $q={1,2,3,4}$ stands for the model used, and $C^{(\rm {est},\mathcal{M}_q)}$ is the estimated class from the $q^{\rm th}$ model. The predictions are made using the relevant predict functions specific to the models deployed. 

\subsection{Localization and Tracking of the Target} \label{Section:local_track}
The $(x,y,z)$ coordinates of the intersection of the laser beams are unique as shown in Fig.~\ref{Fig:detect_new}(a). If one or more laser intersections are blocked by a target, the coordinates of the blocked intersection positions provide the localization of the target at a given time instance. As the target moves, the localization information is also updated in time. The center of the target is the center position of the blocked laser intersections and mesh. For example in Fig.~\ref{Fig:detect_new}(b), the center of the target is at the third blocked intersection of $j=2$ laser mesh. The tracking and mapping of the target's trajectory are obtained as the blocked laser intersection positions are updated as the target moves. Overall SDCLT process of a target is given in Algorithm~\ref{Alg:SDCLM}. In Algorithm~\ref{Alg:SDCLM} at lines $4$ and $5$, a detection test is carried out $S/n_{\rm G} < \gamma$ to determine the presence of a target for each steering position. Once a target is detected, the shape features of the target are extracted (that are provided in Table~\ref{Table:Train_Data}). Next, the target is classified and categorized based on the extracted features. The localization and tracking of the target are also carried out at different steering positions (discussed inSection~\ref{Section:local_track}). The extraction of features, classification, and localization, and tracking of the target are provided at lines $6$ to $9$ of  Algorithm~\ref{Alg:SDCLM}.

\begin{figure*}[!t] 
    \centering
	\begin{subfigure}{0.63\columnwidth}
    \centering
	\includegraphics[width=\textwidth]{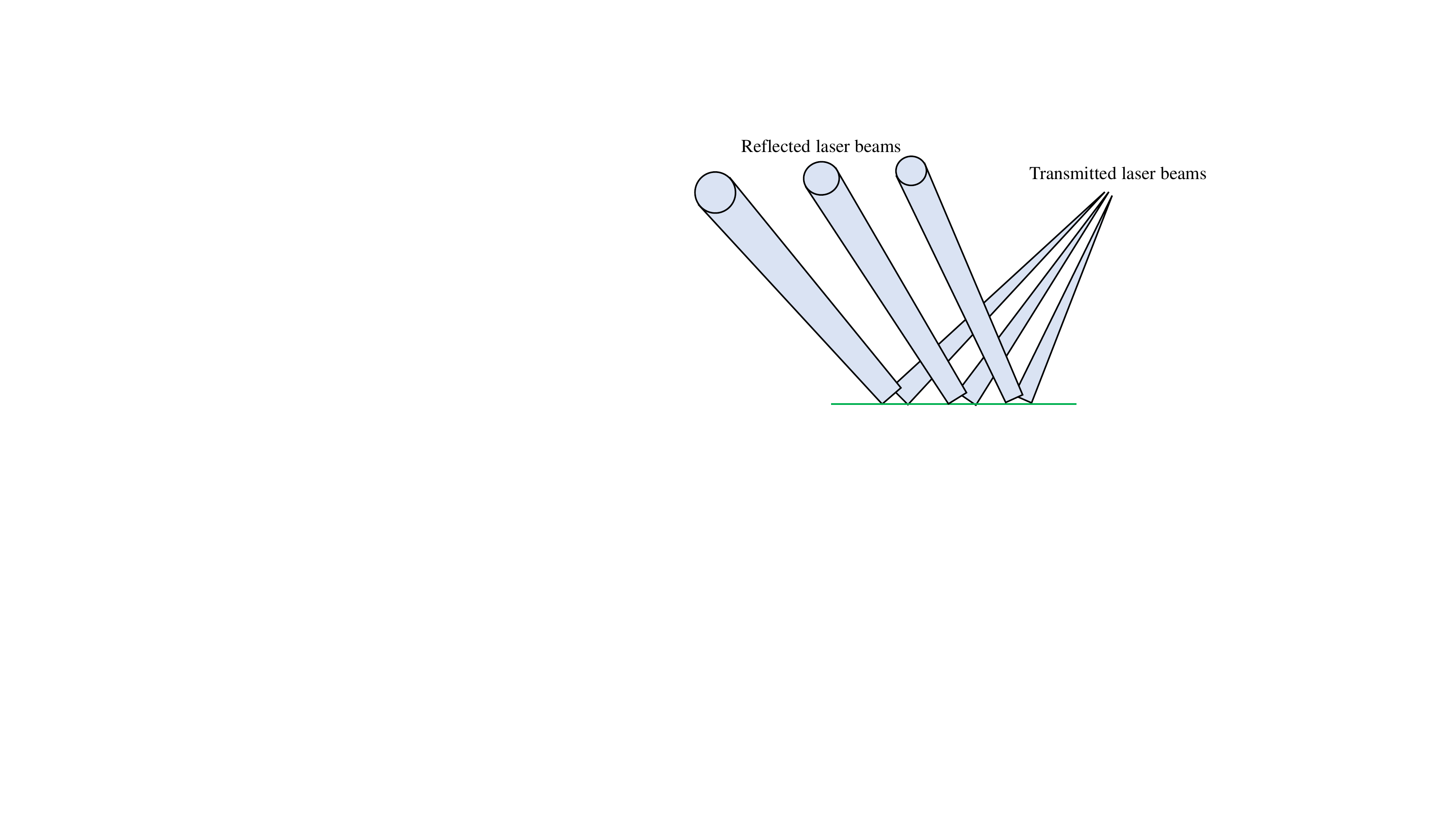}
	  \caption{}  
    \end{subfigure}    
    \begin{subfigure}{0.20\columnwidth}
    \centering
	\includegraphics[width=\textwidth]{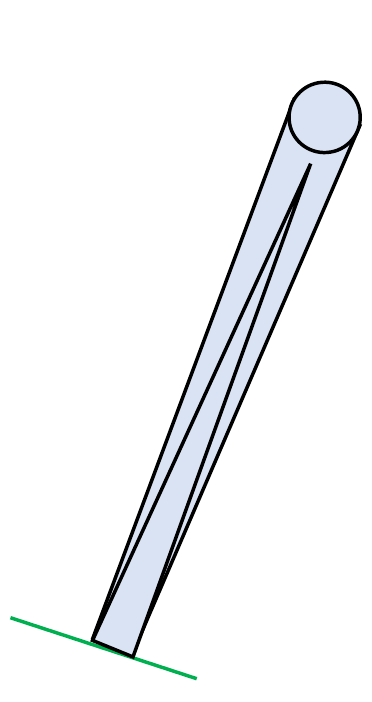}
	  \caption{}  
    \end{subfigure}
    \caption{(a) Reflected laser beams to provide spatial diversity, (b) laser beam transmitted and reflected towards the source.} \label{Fig:mesh_reflect}
\end{figure*}

\begin{figure*}
\centering
\includegraphics[width=0.96\textwidth]{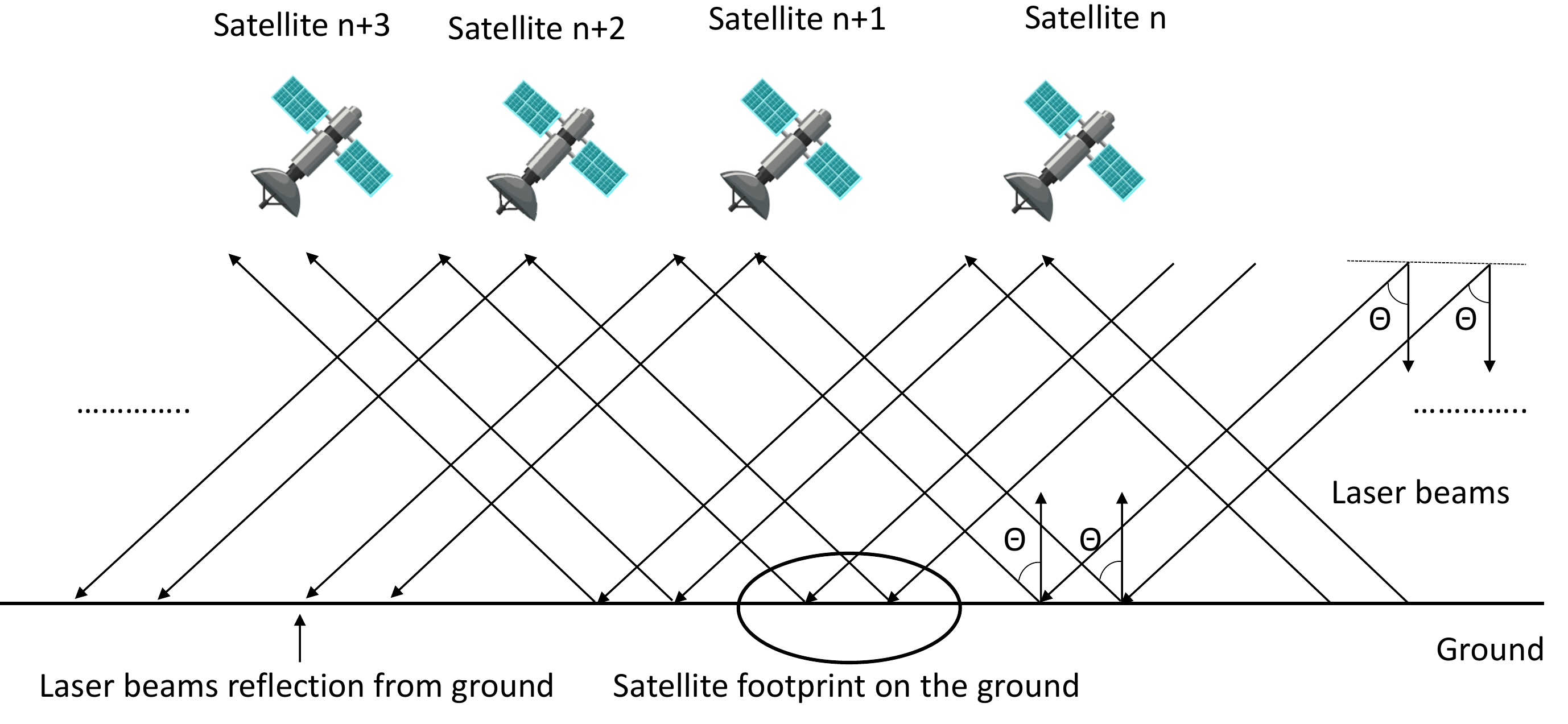}
\caption{An example laser mesh scenario using a low-altitude satellite network. The laser beams from satellites are directed toward the ground at an angle $\Theta$ with the unit normal~(of the satellite). The laser beams are reflected at an angle $\Theta$ with the unit normal~(of the ground) towards the nearby satellite in the network. Using this configuration, and a large number of satellites in the network, the change in the shape of the laser mesh at a given location due to the motion of the satellites can be significantly reduced. }\label{Fig:Sat_exp} 
\end{figure*}

\section{Limitations of the Proposed Approach and Solutions} \label{Section:limitations}
In this section, limitations of our proposed approach are discussed and possible solutions are provided. 

\subsection{Laser Beams Transmission and Reception} \label{Section:Enhancement}

A limitation of the proposed approach is the difficulty in the deployment of airborne laser TXs and a large number of laser RXs on the ground. There can be different options for the deployment of laser TXs and RXs. The TXs of the uniformly spaced array of laser beams can be placed on high-altitude platforms (using tethered hot air balloons) or low-altitude satellites e.g. Starlink satellite network. The RXs of the laser beams can be either located near the TX or co-located with the TX using the reflection of laser beams from the ground as shown in Fig.~\ref{Fig:mesh_reflect}(a), and (b), respectively. This can eliminate the need for the placement of RXs on/near the ground. Similar to Starlink Satellite Constellation, a dense network of satellites can be used carrying TXs and RXs of laser beams shown in Fig.~\ref{Fig:Sat_exp}. In Fig.~\ref{Fig:Sat_exp}, an example laser mesh setup using a network of low-altitude satellites and laser reflection from the ground is shown. The transmitted and reflected laser beams form an angle $\Theta$ with the unit normal at the satellite and ground, respectively. The transmitted laser beams are reflected from the ground and received at a nearby satellite. Using the configuration in Fig.~\ref{Fig:Sat_exp}, and a large number of satellites in the network, the change in the shape of the laser mesh at a given geographic location due to the motion of the satellites is minimum. Moreover, a large coverage area is possible at an affordable cost.

\begin{figure*}[!t]
	\centering
	\includegraphics[width=0.75\columnwidth]{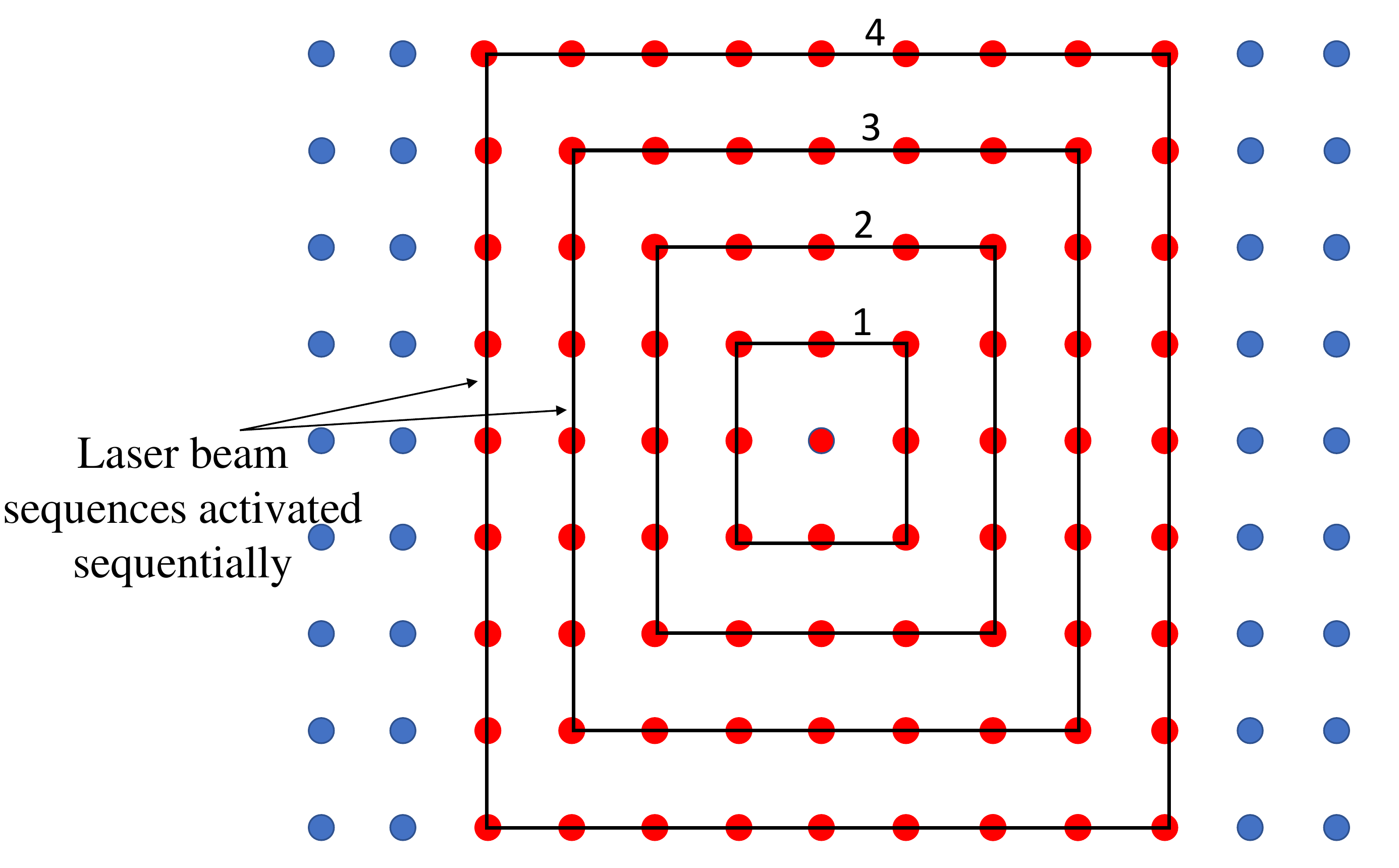} 
	\caption{The sequence of laser beams for multiple target detection. The sequence of laser beams is activated sequentially starting from the center. The center of the sequence is the estimated center of the detected target's position. } \label{Fig:laser_seq} 
\end{figure*}

The classification and localization accuracy of a target depend on the distance between consecutive laser beams in a mesh. To obtain greater accuracy, we require a mesh of closely spaced and spatially diverse laser beams that is challenging. The generation of spatially diverse laser beams can be realized through the use of lens, and mirrors~\cite{thesis}, or using multiple reflectors at different spatial positions in the path of the transmitted laser beams. Fig.~\ref{Fig:mesh_reflect}(a) shows an example scenario of transmitted and reflected laser beams for a given reflector orientation. These techniques, together with the use of multiple hierarchical TXs, and RXs of the laser beams, can create a dense mesh of laser beams that are spatially diverse and in close formation.

\subsection{Multiple Targets Detection} \label{Section:Multiple_targets}
The detection of multiple targets simultaneously is a complex task. In case laser beams are blocked due to multiple targets that are close to each other, a clear decision cannot be made whether it is because of a single large target or multiple smaller targets. To remove this ambiguity, a sequence of laser beams can be directed to the next steering position once a target is identified. The center of the sequences is the estimated center of the target at the next steering position based on the trajectory of the target. The sequences are applied sequentially with a delay of $\Delta t_{\rm seq}$. An example illumination sequence is shown in Fig.~\ref{Fig:laser_seq} where rectangular sequences are applied. In Fig.~\ref{Fig:laser_seq}, $L = 13$, and nine mesh are illuminated sequentially for a delay of $\Delta t_{\rm seq}$ starting from the center outwards for an $i^{\rm th}$ steering position. Other complex sequences that may help in the identification of multiple close flying targets are also possible. The sequential application of the sequences helps in the identification of boundaries of a target, thus leading to the detection of multiple targets.

\begin{figure*}[!t]
	\centering 
	\includegraphics[width=0.72\columnwidth]{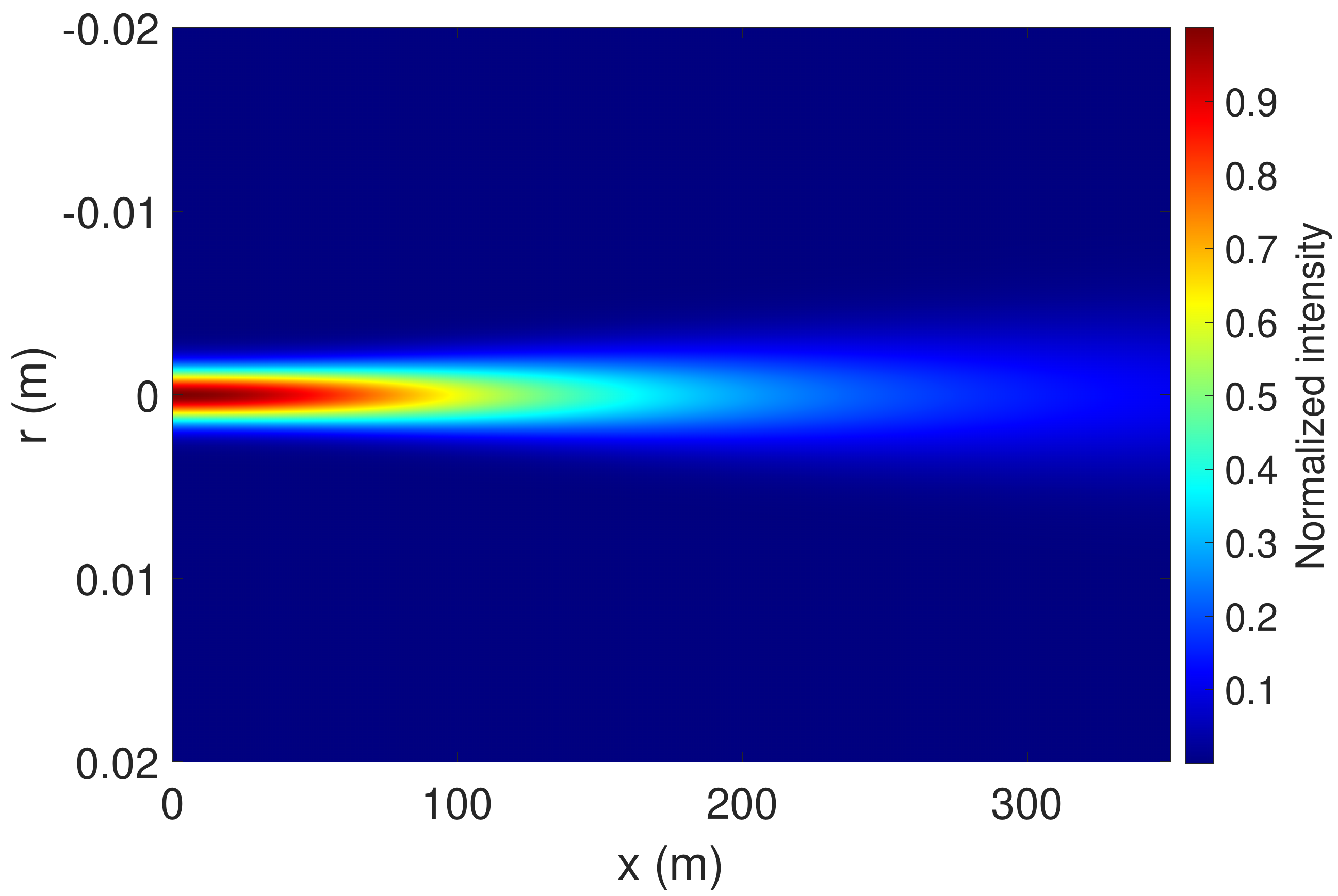} 
	\caption{Intensity plot of the Gaussian laser beam.}
	\label{Fig:Beam_blockage} 
\end{figure*}

\section{Simulation Setup and Results}	\label{Section:Simulation_Results}
Simulations are carried out to first observe the blockage of a laser beam by a target (discussed in Section~\ref{Section:Detection}). In the simulations, a Gaussian beam of waist radius $\num{2e-3}$ is generated as shown in Fig.~\ref{Fig:Beam_blockage}. The intensity of the Gaussian beam shown in Fig.~\ref{Fig:Beam_blockage} is normalized. The wavelength is $\lambda = \num{100e-9}$, beam waist radius $w_0 = \num{2e-3}$, and Rayleigh range $x_{\rm R} = 125.7$~m. An RX is present at a distance of $d_{i,j} = 350$~m, and the radius of the aperture at the RX is $r^{(\rm RX)} = \num{1e-2}$. The variance of the AWGN noise is $\sigma^2_{\rm n} = \num{1e-4}$. The electric field amplitude at the origin is $E_0 = 200$, and the received power in \ref{Eq:RX_pwr} is $\num{3.3e-4}$~W, and the maximum SNR \big(from \ref{Eq:SNR} \big) is $5.1$~dB when there is no target present. The threshold $\gamma$ for a pfa of $0.1$ is set to be $\gamma = 3.6$~dB. Now, a target is introduced at a $200$~m distance from the TX of the laser beam. The value of $\Gamma_1$ and $\Gamma_2$ are selected to be $0.7$ and $0.2$, respectively, for the target. The SNR due to blockage from the target is $-14.7$~dB that is significantly less than the SNR threshold for target detection. Consequently, the target will be detected. 

\begin{figure*}[!t]
	\centering
	\includegraphics[width=0.82\columnwidth]{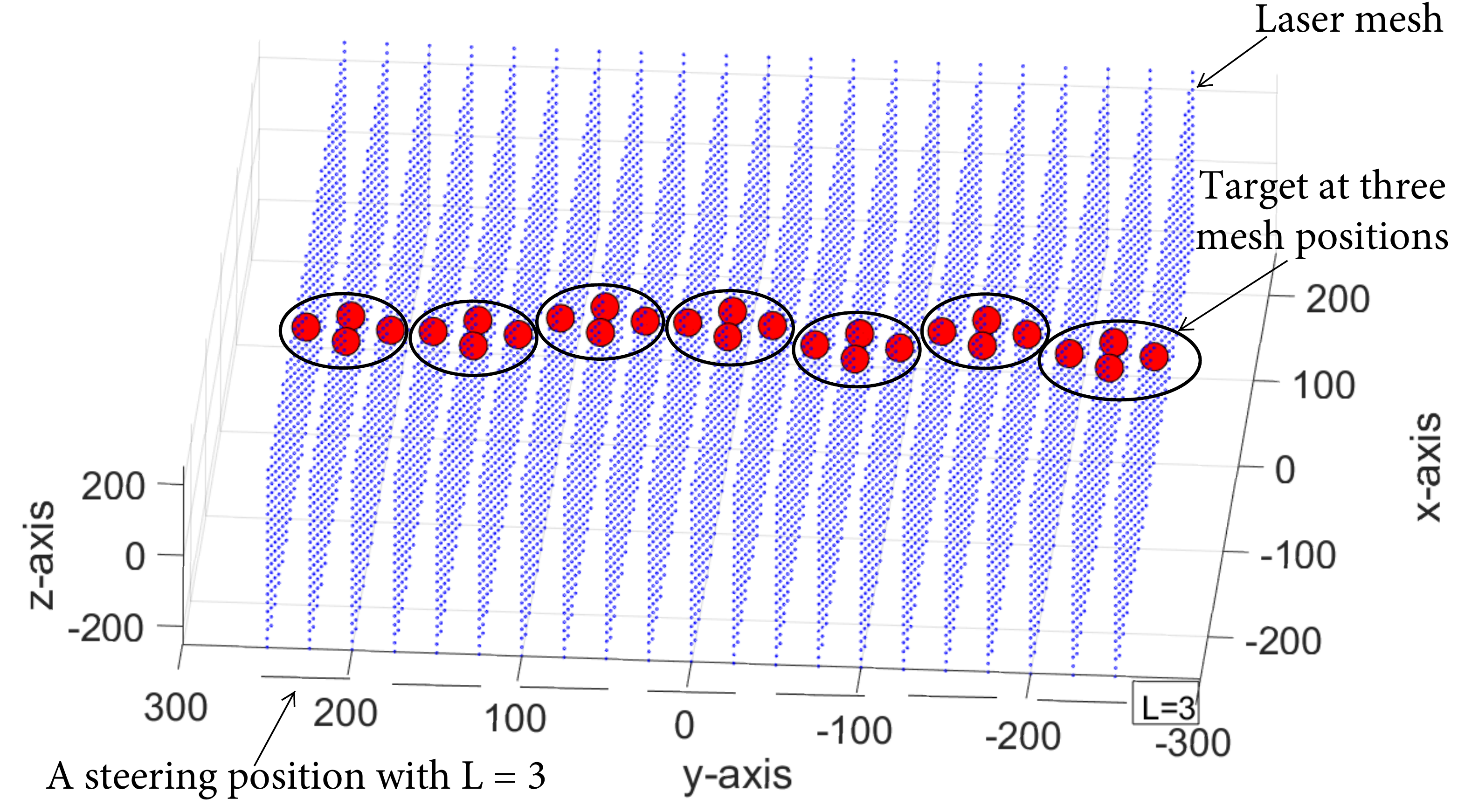} 
	\caption{Simultaneous detection, localization, and tracking of a target using three laser mesh at each steering position. The red dot shows the position of the target. } \label{Fig:target_trajectory} 
\end{figure*}

The laser beam in Fig.~\ref{Fig:Beam_blockage} is used to form meshes at different steering positions shown in Fig.~\ref{Fig:target_trajectory}. In the simulations, we used $7$ steering positions, i.e., $i=-250:25:250$, and each steering position had three 1D arrays i.e. $L=3$. Each 1D array had 21 RX elements from the two airborne UAVs. The number of laser intersection positions in each mesh were $21\times 21$. A target highlighted with red dots is shown in Fig.~\ref{Fig:target_trajectory}. The target lays over three meshes at each steering position. The estimated features of the target from Section~\ref{Section:Features} were recorded for the target. The features of the target are given in Table~\ref{Table:Train_Data} under the name \textit{Given target}. The features of the given target are similar to a BGM-109 Tomahawk missile~\cite{tomahawk}. 

\begin{figure*}[!t] 
    \centering
	\begin{subfigure}{0.47\columnwidth}
    \centering
	\includegraphics[width=\textwidth]{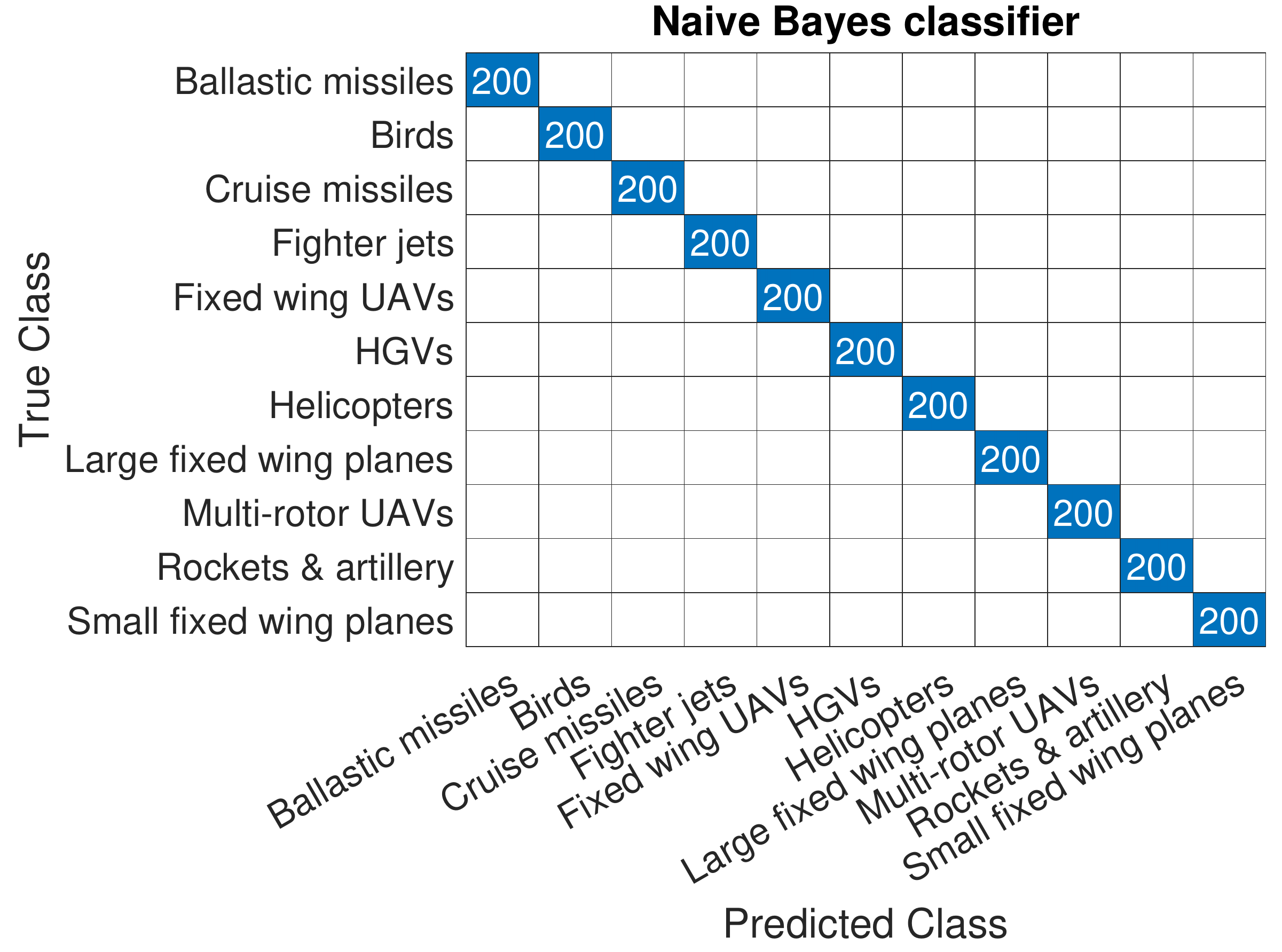}
	  \caption{}  
    \end{subfigure}    
    \begin{subfigure}{0.47\columnwidth}
    \centering
	\includegraphics[width=\textwidth]{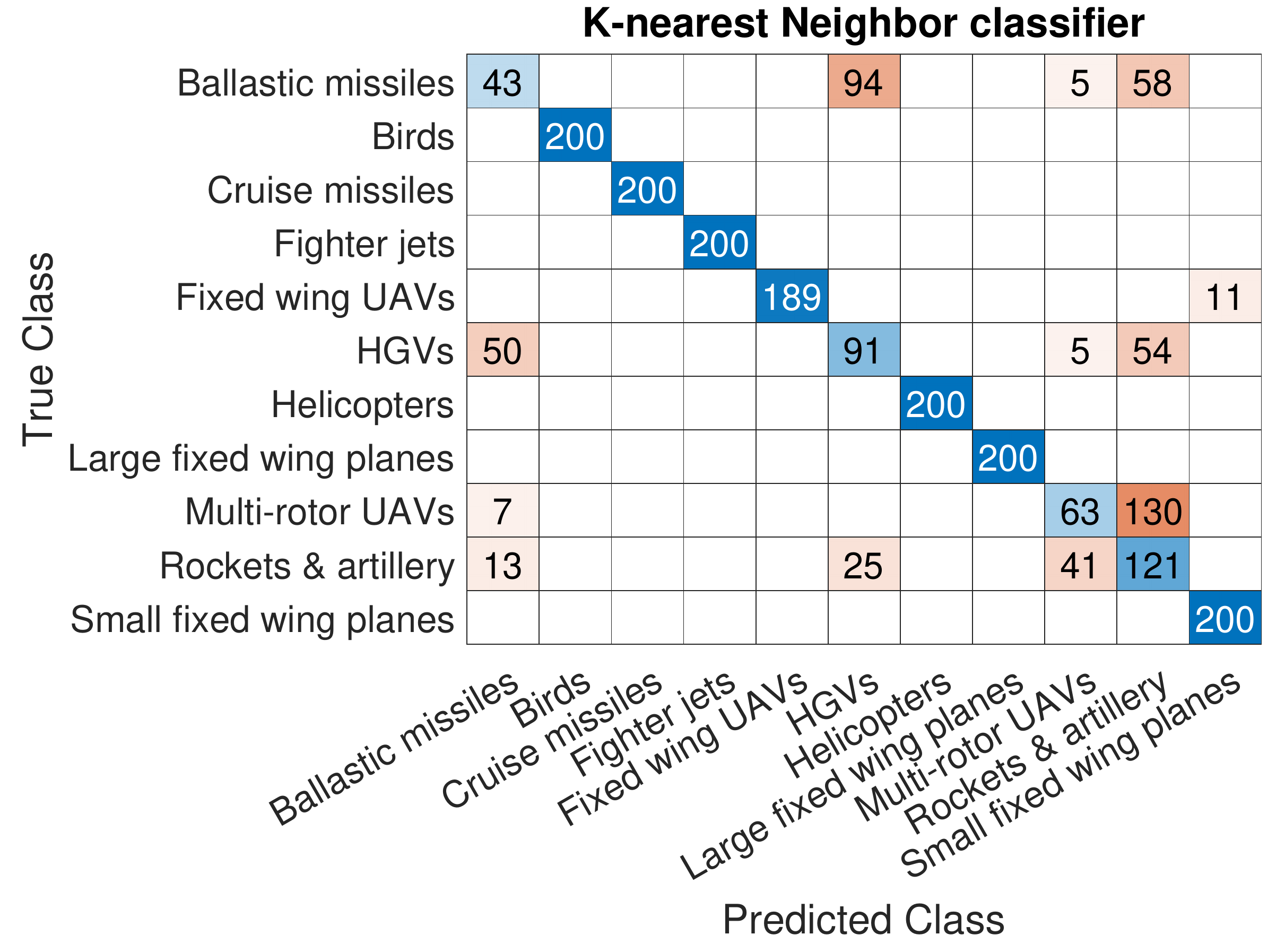}
	  \caption{}  
    \end{subfigure}
\begin{subfigure}{0.47\columnwidth}
    \centering
	\includegraphics[width=\textwidth]{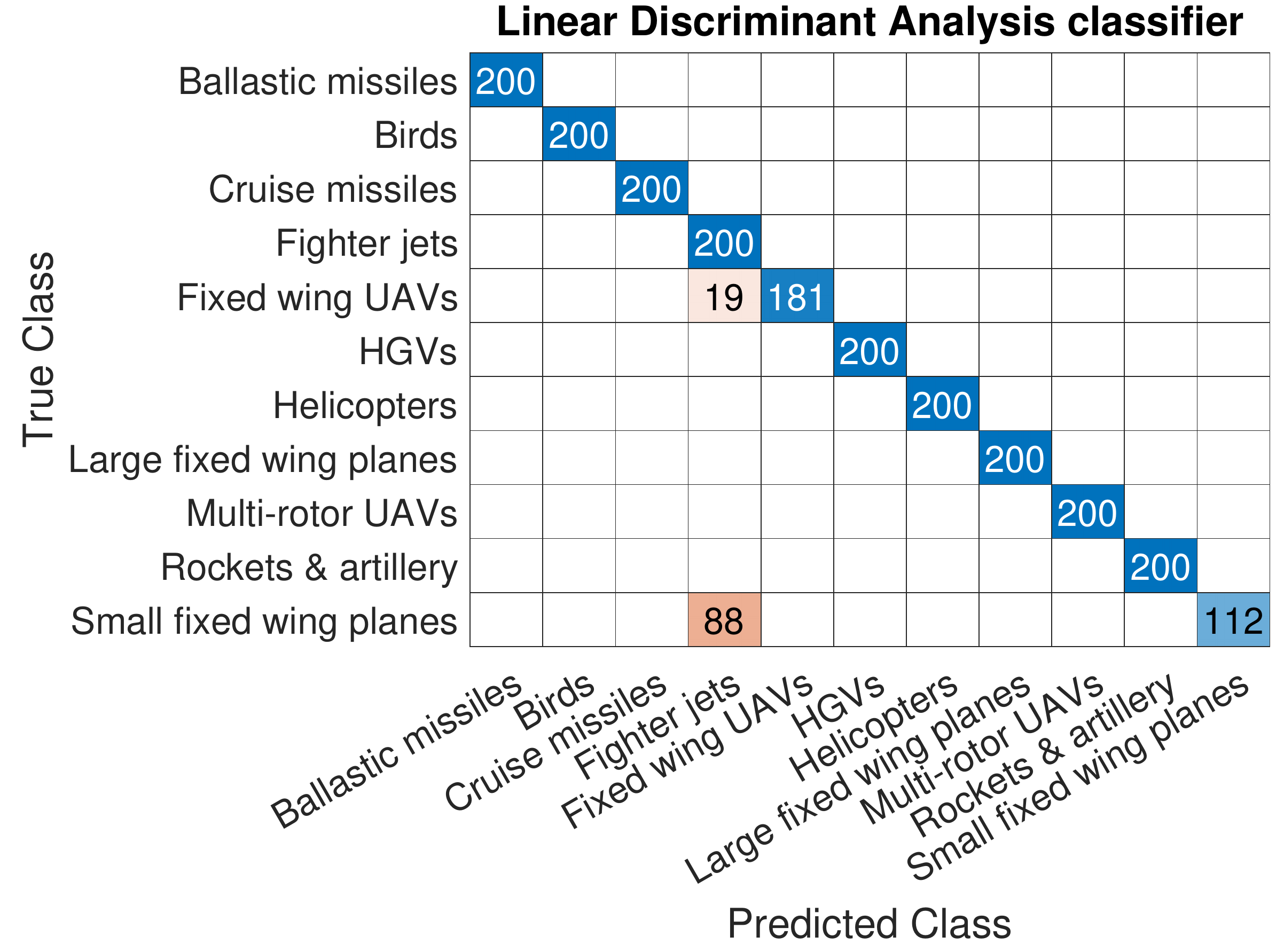}
	  \caption{}  
    \end{subfigure}    
    \begin{subfigure}{0.47\columnwidth}
    \centering
	\includegraphics[width=\textwidth]{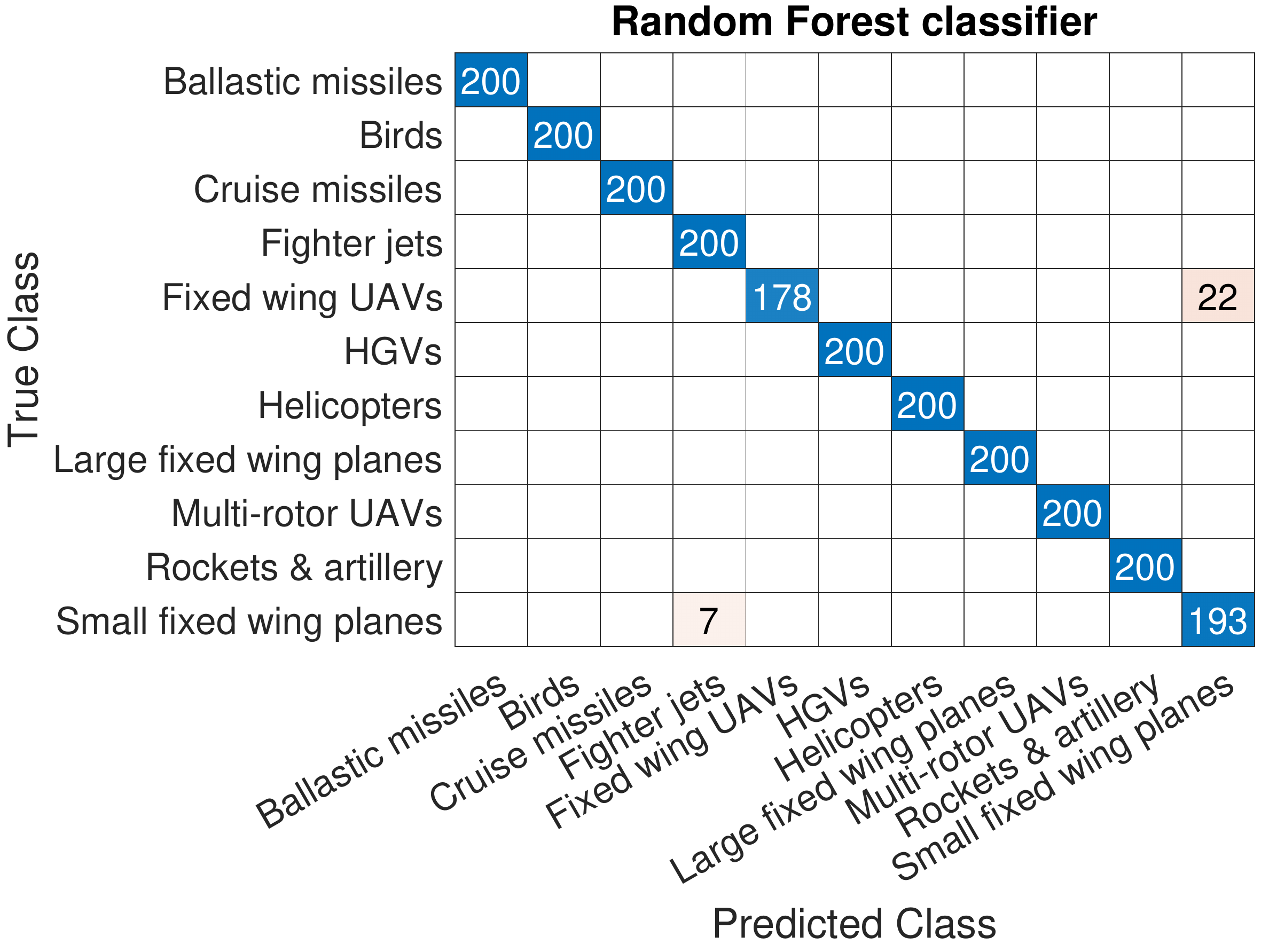}
	  \caption{}  
    \end{subfigure}
    \caption{Confusion matrix for classification of targets at $200$ different instances using (a) Naive Bayes, (b) K-nearest neighbor, (c) Linear Discriminant Analysis classifiers, and (d) Random forest classifiers. The classifications are obtained through automated hyperparameter optimization.} \label{Fig:confusion_mat_nb_lda}
\end{figure*}

In our dataset, we used 200 samples per class generated from the Gaussian distribution parameters given in Table~\ref{Table:Train_Data}. A target can be classified with the help of training data of different aerial targets, and using NB, LDA, KNN, and RF classifiers. The NB, LDA, KNN, and RF classifiers from Statistics and Machine Learning Toolbox of Matlab are used in the simulations~\cite{classifiers}. Hyperparameter optimization in classification is also performed using Matlab. For this specific target (Tomahawk misile), NB, LDA, and DT models were able to correctly classify based on its features provided in Table~\ref{Table:Train_Data} as a cruise missile, whereas KNN failed. For a better understanding of the used classification models, the confusion matrices are also provided for all four classifiers in Fig.~\ref{Fig:confusion_mat_nb_lda}. To derive the confusion matrices, $200$ new samples are created using the parameters given in Table~\ref{Table:Train_Data}. 
The model is optimized using automated optimized hyperparameter values to minimize the classification error. 

The results given in Fig.~\ref{Fig:confusion_mat_nb_lda} show that NB classifier performs the best among all. There are no misclassifications with NB classifier. The NB classifier considers the different features given in Table~\ref{Table:Train_Data} as independent that helps in the best classification. The KNN performs poorly compared to the other classifiers. The classification in KNN is based on nearest distance and values of many of the features of the targets e.g., length and velocity shown in Fig.~\ref{Fig:PDF_len_vel} are overlapping, therefore, KNN misclassifies different targets. The LDA is able to correctly classify most of the targets except fixed-wing UAVs and small fixed-wing planes. The LDA misclassifies fixed-wing UAVs and small fixed-wing planes as fighter jets, as many of the features are similar. Similar to LDA, RF classifier misclassifies fixed-wing UAVs and small fixed-wing planes. However, the total number of misclassifications is smaller for RF classifier compared to the LDA classifier. Overall performance of the models proves the viability of our proposed approach.

\section{Conclusions and Future Work}	\label{Section:Conclusions}
In this work, a novel technique called laser mesh for detection, classification, localization, and tracking of aerial targets that is an alternative to radars is provided. Mesh of laser beams are proposed to detect, classify, and localize aerial targets. To create the mesh, at least two airborne platforms are required. Any aerial object crossing the mesh will block the path of the laser beams and, subsequently, will be detected and localized in space. Using our laser mesh setup, we can obtain the 3D shape, velocity, pitch and drift angles, and a maximum altitude of a target. ML models for classification are used assuming Gaussian distributed features of 3D shape, maximum velocity, and pitch and drift angles, and a maximum altitude of 11 different classes. Simulations proved the viability of the proposed approach. Future work includes carrying out the real-world implementation of the proposed approach.

\section*{Acknowledgement}
This work has been supported by NASA under the Federal Award ID number NNX17AJ94A.

\ifCLASSOPTIONcaptionsoff
  \newpage
\fi

\bibliographystyle{IEEEtran}
\bibliography{References}

\end{document}